\documentclass{emulateapj}

\begin{document}

\def\fluxthres{\hat f_{\bar \e}}
\def\fluxeps{f_{_{\rm \epsilon}}}
\def\zmin{z_{\rm min}}
\def\zmax{z_{\rm max}}
\def\xmin{x_{\rm min}}
\def\xmax{x_{\rm max}}
\def\e{\epsilon}
\def\Estar{{\cal E}_*}
\def\Estarg{{\cal E}_{*\gamma}}
\def\Estargo{{\cal E}_{*\gamma 0}}
\def\Swift{\emph{Swift}}

\slugcomment{\apj, accepted for publication}

\shorttitle{Redshift Distribution of GRBs} \shortauthors{Le \& Dermer}

\title{On the Redshift Distribution of Gamma Ray Bursts in the \Swift~Era}

\author{Truong Le\altaffilmark{1} and Charles D. Dermer\altaffilmark{2}}
\affil{E. O. Hulburt Center for Space Research \break
Naval Research Laboratory, \break
Washington, DC 20375, USA}

\altaffiltext{1}{truong.le@nrl.navy.mil}
\altaffiltext{2}{dermer@gamma.nrl.navy.mil}

\begin{abstract}

A simple physical model for long-duration gamma ray bursts (GRBs)
is used to fit the redshift ($z$) and the jet opening-angle
distributions measured with earlier GRB missions and with \Swift.
The effect of different sensitivities for GRB triggering is
sufficient to explain the difference in the $z$ distributions of
the pre-\Swift~and \Swift~samples, with mean redshifts of $\langle
z \rangle \cong 1.5$ and $\langle z \rangle \cong 2.7$,
respectively. Assuming that the emission properties of GRBs do not
change with time, we find that the data can only be fitted if the
comoving rate-density of GRB sources exhibits positive evolution
to $z \gtrsim 3$ -- 5.  The mean intrinsic beaming factor of GRBs
is found to range from $\approx 34$ -- 42, with the \Swift~average
opening half-angle $\langle \theta_j\rangle\sim 10^\circ$,
compared to the pre-\Swift~average of $\langle \theta_j\rangle\sim
7^\circ$. Within the uniform jet model, the GRB luminosity
function is $\propto L^{-3.25}_*$, as inferred from our best fit
to the opening angle distribution. Because of the unlikely
detection of several GRBs with $z \lesssim 0.25$, our analysis
indicates that low redshift GRBs represent a different population
of GRBs than those detected at higher redshifts. Neglecting
possible metallicity effects on GRB host galaxies, we find that
$\approx 1$ GRB occurs every 600,000 yrs in a local $L_*$ spiral
galaxy like the Milky Way. The fraction of high-redshift GRBs is
estimated at 8 -- 12\% and 2.5 -- 6\% at $z \geq 5$
 and $z\geq 7$, respectively, assuming continued positive
evolution of the GRB rate density to high redshifts.
\end{abstract}


\section{INTRODUCTION}

Gamma-ray bursts (GRBs) are brief flashes of $\gamma$ rays
occurring at an average rate of a few per day throughout the
universe. The ultimate energy source of a GRB is believed to be
associated with a catastrophic event associated with black-hole
formation~\citep{mr97}. This event takes place through the
collapse of the core of a massive star in the case of
long-duration GRBs, and due to merger or accretion-induced
collapse events for the short-hard class of GRBs. For the class of
long-duration GRBs, which is the only type considered in this
paper, liberation of energy from the GRB source is thought to result
in a very high temperature, baryon-dilute fireball that expands
to reach highly relativistic speeds and form a collimated outflow.
Dissipation of the directed kinetic energy of the relativistic
blast wave through internal and external shocks is the standard
model for producing the prompt radiation and afterglow
\citep[see][for a recent review]{mes06}.

Evidence of jetted GRBs is obtained from radio~\citep{wkf98} and
optical~\citep{sta99} observations of achromatic breaks in the
afterglow light curves. The structure of these jets is, however,
still an open question. The two leading models are (1) the uniform
jet model, in which the energy per solid angle $\partial {\cal
E}_*/\partial \Omega$ is roughly constant within a well-defined
jet opening angle $\theta_j$, but drops sharply outside of
$\theta_j$, with $\theta_j$ differing between
GRBs~\citep[see][]{mrw98, mr97}; and (2) the universal structured
jet model~\citep{rlr02,zha04}, in which all GRB jets are
intrinsically identical with the directional energy release
$\partial\Estar/\partial \Omega$ dropping approximately as the
inverse square of the opening angle from the jet axis.

\citet{fra01} and~\citet{pk01} found that the absolute
$\gamma$-ray energy releases, after correcting for the jet opening
angles inferred from the afterglow light curves, are clustered
near $\Estarg \cong 5\times 10^{50}$ ergs. \citet{blo03}, treating
a larger sample, confirmed this clustering around $\Estarg \sim
1.3 \times 10^{51} \ \rm ergs$ (the total energy release $\Estar =
\oint d\Omega \,(\partial \Estar/\partial \Omega)
> \Estarg$ depends on the assumed efficiency for $\gamma$-ray energy conversion).
In a later analysis by \citet{fb05}, still in the pre-\Swift~era,
a broader spread by about one order of magnitude is found in the
beaming corrected $\gamma$-ray energy release around a mean value
of $\Estarg\approx 2\times 10^{51}$ ergs (see Fig.\ 1). Since we
use jet opening angles derived from these analyses, which are
based on the uniform jet model, we also adopt this model in our
treatment. The sample of \citet{fb05} is taken to provide a good
representation of the pre-\Swift~observations of GRBs.
\begin{figure}[htp]
\vskip+0.1in \hskip-0.46in
\includegraphics[width=3.8in]{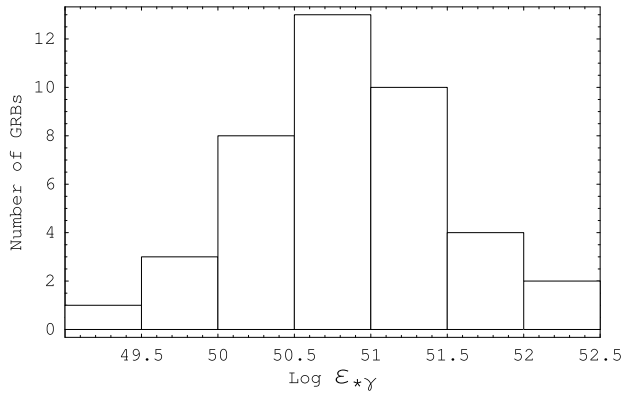}
\vskip-2.28in

\caption{Histogram of the corrected absolute $\gamma$-ray energy
release $\Estarg$ for pre-\Swift~GRBs \citep{fb05}.}

\label{fig1}
\end{figure}
GRBs can in principle exist \citep[e.g.,][]{sch99,mm96} and be
detected to very high redshift, $z\gtrsim 10$ \citep{cl00,lr00},
which holds promise that GRBs can be used to probe the early
universe. Their utility for this purpose depends on the frequency
and brightness of high-redshift GRBs, which in turn depends on the
rate-density evolution of GRB sources and the evolution of GRB
source properties with cosmic time. Given the assumption that the
GRB rate density is proportional to the measured star formation
rate (SFR) history of the universe~\citep{tot99,bn00}, the
redshift distribution of GRBs can be predicted \citep{bl02}.
Correlations between spectral properties and energy releases may
also permit GRBs to be used to determine cosmological
parameters~\citep[e.g.,][]{ghi04,ms05}.
\begin{figure}[htp]
\vskip+0.1in \hskip-0.46in
\includegraphics[width=3.8in]{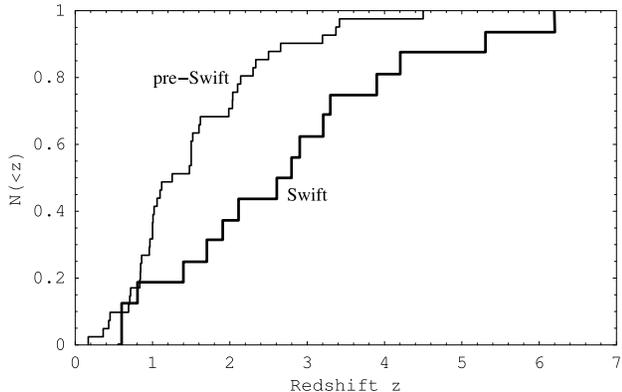}
\vskip-2.21in

\caption{The cumulative redshift distribution of GRBs for 41 GRBs
in the pre-\Swift~sample~\citep[thin histograms;][]{fb05}, and 16
GRBs in the \Swift~sample~\citep[thick histograms;][]{jak06}. The
median redshifts of the pre-\Swift~and \Swift~bursts are $z
\approx 1.25$ and $ z \approx 2.70$, respectively.}

\label{fig2}
\end{figure}

With the launch of the \Swift~satellite~\citep{geh04}, rapid
follow-up studies of GRBs triggered by the Burst Alert Telescope
(BAT) on \Swift~became possible. A fainter and more distant
population of GRBs than found with the pre-\Swift~satellites
CGRO-BATSE, BeppoSAX, INTEGRAL, and HETE-2 is
detected~\citep{ber05}. The mean redshift of 41 pre-\Swift~GRBs
that also have measured beaming breaks~\citep{fb05} is $\langle
z\rangle = 1.5$, while 16 GRBs discovered by \Swift~have $\langle
z\rangle = 2.72$~\citep{jak06} (see Fig.~\ref{fig2}).
\citet{bag06} performed five different statistical tests to
compare the redshift distributions of \Swift~and pre-\Swift~GRBs
assuming, as the null hypothesis, that they are identical
distributions. They conclude that the redshift distributions are
different, that is, all five tests reject the null hypothesis at
the $\geq 97\%$ significance level. Thus, the pre-\Swift~and
\Swift~data sample distinct subsets of the overall GRB population
due to different detector characteristics and followup
capabilities.

Here, we develop a physical model to understand the differences
between the redshift distributions of \Swift~and pre-\Swift~GRBs,
taking into account the different detector triggering properties.
The GRB model is based on the uniform jet model for GRBs, though
the typical properties are assigned in accordance with GRB
observations. The GRB rate density and distribution of jet opening
angles are the principle unknowns, with parameter values for the
mean intrinsic duration and absolute $\gamma$-ray energy release
of GRBs determined from observations and best fits to the data. A
GRB is detected if its flux exceeds the flux threshold within the
energy window of a detector \citep[the actual detection criteria
are more complicated; see][and discussion below]{ban03,ban06}. For
a given rate density of GRBs, the redshift and opening angle
distributions are calculated and fit to the data.

We consider only data for the long, soft population of GRBs. The
separation of the long duration GRBs from the short, hard GRBs,
which clearly comprise a separate source population is complicated
both by the extended emission observed with Swift from GRBs that
would normally be classified as short hard GRBs
\citep[e.g.,][]{bar05}, and the appearance of the unusual GRB
060614 that defies simple classification
\citep{geh06,gal06,fyn06}. We therefore additionally assume that
the data set used to test our model represents only, or mostly,
classical long duration GRBs. We return to this point in the
discussion.

In \S~2, we develop the formulation for bursting sources, and we
fit the data in \S 3. Inferences for the differences between the
\Swift~and pre-\Swift~redshift distributions are also made in
\S~3, including implications for beaming factors and low-$z$ GRBs.
We conclude in \S~4 by discussing the derived GRB source rate
density, the nature of the low-redshift GRBs with weak energies,
and the predicted fraction of high-redhift GRBs. Derivation of the
bursting rate of sources in a flat $\Lambda$ cold dark matter
cosmology is given in the Appendix.

\section{Statistics of Cosmological GRBs}

\subsection{Model GRB}

We approximate the spectral and temporal profiles of a
GRB occurring at redshift $z$ by an emission spectrum that is constant
for observing angles
$\theta = \arccos\mu \leq \theta_j = \arccos
\mu_j$ to the jet axis during the period $\Delta t_*$; thus the
duration of the GRB in the observer's frame is simply $\Delta t =
(1+z) \Delta t_*$ (stars refer to the stationary frame, and terms
without stars refer to observer quantities).  The $\nu F_\nu$ spectrum
is therefore written as
\begin{equation}
\nu F_\nu \equiv \fluxeps(t) \cong f_{\epsilon_{pk}} S(x)
H(\mu; \mu_{j},1) \ H(t;0, \Delta t) \ , \\
\label{eq1}
\end{equation}
where the spectral function $S(x) = 1$ at $x = \e/\e_{pk} =
\e_*/\e_{pk*}$.
At observer time $t$, $\epsilon = h\nu/m_e c^2$ is the
dimensionless energy of a photon in units of the electron
rest-mass energy, and $\epsilon_{pk}$ is the photon energy at
which the energy flux $f_\epsilon $ takes its maximum value
$f_{\epsilon_{pk}}$.  The quantity $H(\mu; \mu_{j}, 1)$ is the
Heaviside function such that $H(\mu; \mu_{j}, 1) = 1$ when $\mu_j
\leq \mu \leq 1$, that is, the angle $\theta $ of the
observer with respect to the jet axis is within the opening angle
of the jet, and $H(\mu; \mu_{j}, 1) = 0$ otherwise.  One possible
approximation to the GRB spectrum is a broken power law, so that
\begin{equation}
S(x) = \left[ \left(\frac{\epsilon}{\epsilon_{pk}}\right)^a
H(\epsilon_{pk} - \epsilon) +
\left(\frac{\epsilon}{\epsilon_{pk}}\right)^b H(\epsilon -
\epsilon_{pk}) \right] \ , \\
\label{Lambda}
\end{equation}
where the Heaviside function $H(u)$ of a single index is defined
such that $H(u) = 1$ when $u\geq 0$ and $H(u) = 0 $ otherwise. The
$\nu F_\nu$ spectral indices are denoted by $a(>0)$ and $b(<0)$.
Note that $S(x)$ can easily be generalized to accommodate spectral
energy distributions (SEDs) with multiple components.

The bolometric fluence of the model GRB for observers with $\theta
\leq \theta_j$ is given by
\begin{equation}
F = \int^\infty_{-\infty} dt \; \int^\infty_0 d\e\; \frac{\fluxeps(t)
}{\epsilon}  = \lambda_b\; f_{\epsilon_{pk}}\;\Delta t  \;,
\label{eq2}
\end{equation}
where $\lambda_b$ is a bolometric correction to the peak measured $\nu
F_\nu$ flux. If the SED is described by eq.\ (\ref{Lambda}), then
$\lambda_b = (a^{-1} - b^{-1})$, and is independent of $\e_{pk}$. The
beaming-corrected $\gamma$-ray energy release $\Estarg$ for a
two-sided jet is given by
\begin{equation}
\Estarg = 4 \pi d^2_L (1-\mu_j) \; \frac{F}{1+z}  \ , \\
\label{eq3}
\end{equation}
where the luminosity distance
\begin{equation}
d_L(z) = \frac{c}{H_0}(1+z) \ \int^z_0
\frac{dz^\prime}{\sqrt{\Omega_m (1+z^\prime)^3 + \Omega_\Lambda}}
\label{eq4}
\end{equation}
for a $\Lambda$CDM universe.  Substituting eq.~(\ref{eq2}) for $F$
into eq.~(\ref{eq3}) gives the peak flux
\begin{equation}
f_{\epsilon_{pk}} = \frac{\Estarg}{4 \pi d^2_L(z) (1 - \mu_j)
\Delta t_* \ \lambda_b} \ . \\
\label{eq5}
\end{equation}
Finally, by substituting eq.\ (\ref{eq5}) into eq.\ (\ref{eq1}),
the energy flux becomes
\begin{eqnarray}
\fluxeps(t)  \; = \; \frac{\Estarg \;H(\mu; \mu_j,1) \ H(t;0,
\Delta t) S(x)}{4 \pi d^2_L(z) (1 - \mu_j) \Delta t_* \lambda_b}
\;. \label{eq6}
\end{eqnarray}
Because of the reduced spectral sensitivity of Swift to high-energy
($\gtrsim 200$ keV) emission and the measurement of $\e_{pk}$ in many GRBs, and
to avoid complications from the potentially large number
of parameters in our GRB model, we consider
in this paper the simplest GRB SED with $a = b = 0$, so that
$\lambda_b \rightarrow \ln ({x_{\rm max}/x_{\rm min}})$ (see
eqs.~[\ref{eq2}], [\ref{eq5}] and [\ref{eq6}]). The flat part
of the spectrum takes the value of peak flux $f_{\epsilon_{pk}}$ given
by eq.\ (\ref{eq5}), and is
assumed to intersect the sensitive range of the GRB detector.
By making the above approximation, we modify the fractional number
of bursts that occurs within a given redshift $z$ and within a jet
opening angle $\theta_j$ (see eqs.~[\ref{eq18}], [\ref{eq19}], and
[\ref{eq20}]); however, the physical interpretation of the problem
is expected not to change drastically when $a, b \neq 0$, which we
plan to address in a subsequent paper. In the flat spectrum
approximation, a $\nu F_\nu$ spectrum ($a = b = 0$) has to be cut
off at low ($x_{\rm min}$) and high ($x_{\rm max}$) energies in
order to avoid divergent flux. The model spectrum is not likely to
extend over more than two orders of magnitude, so that $ \lambda_b
\lesssim 5$. Thus we take $\lambda_b$, the bolometric correction
factor, equal to $5$ in our calculations.

\subsection{Bursting Rate of GRB Sources}

To calculate the redshift and the jet opening angle distributions,
we need to determine the rate at which GRBs occur per steradian.
For a $\Lambda CDM$ cosmology with $\Omega_m = 0.27$,
$\Omega_\Lambda = 0.73$, and $H_0 = 72 \; \rm ~km \ s^{-1} \
Mpc^{-1}$ \citep{spe03}, the event rate per unit time per sr for
bursting sources is
\begin{eqnarray}
\frac{d\dot{N}}{d\Omega}\; & = &\; \frac{c}{H_0} d^2_L(z) \
\dot{n}_{co}(\Estarg,\mu_j,\epsilon_{pk*},\Delta
t_*,z) \nonumber \\
& \times & \frac{\ d\Estarg \ d\mu_j \ d\mu \ d\epsilon_{pk*} \
d(\Delta t_*) \ dz}{(1+z)^3 \sqrt{\Omega_m (1+z)^3 +
\Omega_\Lambda}} \ , \label{eq8}
\end{eqnarray}
(derivation of this result is given in Appendix A).  Here,
$\dot{n}_{co}(\Estarg,\mu_j,\epsilon_{pk*},\Delta t_*,z) =
\dot{n}_{co}(z)$ $ Y(\Estarg,\mu_j,\epsilon_{pk*},\Delta t_*)$ is
the distribution of GRB sources differential in $\Estarg,$
$\mu_j,$ $\epsilon_{pk*},$ $\Delta t_*,$ and $z$, and separability
of the comoving rate density $\dot{n}_{co}(z) = \dot{n}_{co}
\Sigma_{_{SFR}}(z)$ of GRBs from their properties is assumed,
where $\Sigma_{_{SFR}}(z)$ is GRB star formation rate function.
This is the crucial assumption of this treatment, and so needs
restated. There is assumed to be no evolution of the emission
properties of GRB with cosmic time. Thus, for example, the mean
jet opening angles, the mean energies, and the mean values of
$\e_{pk*}$ (which could be correlated with $\Estarg$ to satisfy
the \citet{ama02} and \citet{ghi04} relations when $a, b \neq 0$)
do not change through cosmic time. The phenomenological validity
of the the Amati and Ghirlanda relations for GRBs at different
values of $z$ suggests the plausibility of this assumption. These
relations can be introduced in more detailed treatments by
considering source spectral properties where $\e_{pk*}$ is
correlated with $\Estarg$.

Employing an integral formulation for beamed
sources~\citep[2006]{der92}, the observed directional event rate for
bursting sources\footnote{In a more accurate treatment of detector
response, one should calculate a photon-energy integration over the
energy-dependent effective area, rather than describing a $\gamma$-ray
telescope by a $\nu F_\nu$ flux sensitivity $\fluxthres$ by the mean
photon energy $\epsilon$ of the sensitive waveband of the detector. A
further complication, not considered here, is that $\fluxthres$ may
depend on the duration $\Delta t$ of the GRB.}  with $\nu F_\nu$
spectral flux greater than $\fluxthres$ at observed photon energy $\e$
is given by
$$
\frac{d\dot{N}(> \fluxthres)}{d\Omega} = \frac{c}{H_0}
\int^\infty_{\fluxthres} df^{'}_{\epsilon} \int^\infty_0 d\Estarg
\int^\infty_0 d\epsilon_{pk*} \int^\infty_0 d(\Delta t_*)
$$

$$
\times \int^{\mu_{\rm jmax}}_{\mu_{\rm jmin}} d\mu_j
\int^1_{\mu_j} d\mu \int^\infty_0 dz\;\frac{d^2_L(z)
\dot{n}_{co}(z)Y(\Estarg,\mu_j,\epsilon_{pk*},\Delta t_*)
}{(1+z)^3 \ \sqrt{\Omega_m (1+z)^3 + \Omega_\Lambda}} $$

\vspace{+2mm}
\begin{equation}
\hspace{-31mm} \times \; \; \delta \left[f^{'}_{\epsilon} -
f_{\epsilon} (z, \Estarg,\mu_j,\epsilon_{pk*},\Delta t_*) \right]
. \label{eq9}
\end{equation}
This expression can also be written in terms of a Heaviside
function involving a minimum flux threshold condition $f_\e \geq
\fluxthres$. For discrete values of $\Estarg$, $\e_{pk*}$, and
$\Delta t_*$,
\begin{eqnarray}
Y(\Estarg,\mu_j,\epsilon_{pk*},\Delta t_*) \; & = & \; g(\mu_j) \
\delta(\Estarg- \Estargo) \; \delta(\epsilon_{pk*} -
\epsilon_{pk{*0}}) \nonumber \\
& \times & \; \delta(\Delta t_* - \Delta t_{*0}) \ , \label{eq10}
\end{eqnarray}
where $g(\mu_j)$ is the jet opening angle distribution, and
$\fluxthres$ is the instrument's detector sensitivity. The
sensitivity of a detector to a GRB depends on many factors,
namely, the detector area, the detector efficiency, the fraction
of the detector that is active, the fraction of the coded mask
that is open, the average solid angle, and the internal
background~\citep{ban03}.  \citet{geh04} give the BAT's instrument
detector sensitivity to be about $\sim 10^{-8} \ \rm ergs \
cm^{-2} \ s^{-1}$, which is about an order of magnitude more
sensitive than BATSE and Beppo-SAX, particularly in the case of
those GRBs for which redshifts were obtained, as discussed below.

We can make a rough estimate of the detector threshold for both
\Swift~and pre-\Swift~instruments, noting that the mean flare size
$\langle \phi \rangle = u\phi_{\rm
min}/(u-1)$ for a power law
size distribution with index $u~ (>1; ~u=3/2$ for a Euclidean
distribution). For Swift GRBs, we took the ratio of the measured
GRB fluences to their $\langle t_{90} \rangle$
durations\footnote{swift.gsfc.nasa.gov} to find that the mean flux
of a detected Swift GRB is $\sim 5 \times 10^{-8}  \ \rm ergs\
cm^{-2} \ s^{-1}$.  For an index $u \cong 1.1$ -- 1.2, as
appropriate to the Swift range, we find that the Swift is
sensitive to GRBs as weak as 10$^{-8}  \ \rm ergs \ cm^{-2} \
s^{-1}$. We take this energy flux as the effective Swift flux
threshold, recognizing that the actual detection of GRBs with
Swift is more complicated, involving both a rate and image trigger
\citep{ban06}.

The assignment of an effective flux threshold for the
pre-\Swift~detectors is complicated by the fact that they involve
a variety of detectors---BATSE, Beppo-SAX, HETE-II, and
INTEGRAL---each with its own specific triggering criteria. The
mean flux of the pre-\Swift~ sample can be obtained from the
values of fluence and $t_{90}$ compiled by \citet{fb05}. We find
that the mean flux of the pre-\Swift~GRBs is $\sim 7 \times
10^{-7} \ \rm ergs \ cm^{-2} \ s^{-1}$, meaning that the flux
threshold sensitivity is a factor $\approx 5$ less than this
value, because $u\cong 1.2$ at fluxes $\gtrsim 10^{-7}  \ \rm
ergs\ cm^{-2} \ s^{-1}$ \citep{pac99}. For simplicity, we use the
value of  10$ ^{-7} \ \rm ergs \ cm^{-2} \ s^{-1}$ for the mean
sensitivities of the GRB detectors prior to Swift. Hence, in this
work we assume the detector threshold for \Swift~and
pre-\Swift~GRBs to be $\sim 10^{-8}$ and $\sim 10^{-7} \ \rm ergs
\ cm^{-2} \ s^{-1}$, respectively.

Integrating over $\mu$ in eq.~(\ref{eq9}) gives the factor $( 1-
\mu_j)$ describing the rate reduction due to the finite jet
opening angle. Hence, eq.~(\ref{eq9}) becomes
$$
\frac{d\dot{N}(>\fluxthres)}{d\Omega} = \frac{c}{H_0}
\int^\infty_{\fluxthres} df^{'}_{\epsilon} \int^{\mu_{\rm
jmax}}_{\mu_{\rm jmin}}d\mu_j \; g(\mu_j)(1-\mu_j)
$$

\begin{equation}
\times \; \int^\infty_0 dz\; \frac{d^2_L(z) \ \dot{n}_{co}(z) \
\delta (f^{'}_{\epsilon} - f_{\epsilon})}{(1+z)^3 \ \sqrt{\Omega_m
(1+z)^3 + \Omega_\Lambda}} \;, \label{eq11}
\end{equation}
where $f_\epsilon$ is given by eq.~(\ref{eq6}).

In order to close the system of equations and solve for the redshift
distribution, we must also specify the functional form of $g(\mu_j)$
as well as the comoving GRBs rate density $\dot{n}_{co}(z)$. GRBs with
small opening angles, that is, with $\theta_j \ll 1$, will radiate
their available energy into a small cone, so that such GRBs are
potentially detectable from larger distances (though not always, as a
different portion of the spectrum is observed), with their rate
reduced by the factor $(1-\mu_j)$. By contrast, GRB jets with large
opening angles are more frequent but only detectable from
comparatively small distances. The intrinsic opening angle distribution
can be much different than the measured opening-angle distribution
because of these effects~\citep{gpw05}.

Since the form for the jet opening angle $g(\mu_j)$ is unknown, we
consider the function
\begin{equation}
g(\mu_j) = g_0 \ (1-\mu_j)^s \ H(\mu_j;\mu_{\rm jmin},\mu_{\rm
jmax}) \;,
\label{eq12}
\end{equation}
where $s$ is the jet opening angle power-law index; for a
two-sided jet, $\mu_{\rm jmin} \geq 0$. Because the distributions
are normalized to unity,
\begin{equation}
g_0 = \frac{1+s}{(1-\mu_{\rm jmin})^{1+s} - (1-\mu_{\rm
jmax})^{1+s}} \;.
\label{eq13}
\end{equation}
\begin{figure}[]
\vskip+0.1in \hskip-0.46in
\includegraphics[width=3.8in]{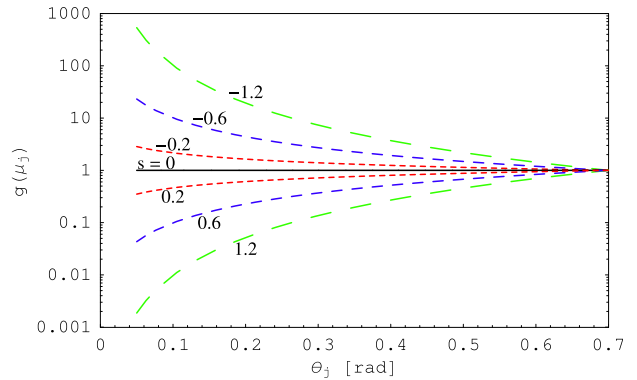}
\vskip-2.2in

\caption{The number of sources per unit $\theta_j$ angle for $s =
-1.2$, $-0.6$, $-0.2$, $0$, $0.2$, $0.6$, $1.2$ in
eq.~(\ref{eq12}) between $\theta_j = 0.05$ and $0.7$ radians. }
\label{fig3} \vskip+0.1in
\end{figure}
Equation~(\ref{eq12}) hardly exhausts the possible expressions for
$g(\mu_j)$, but contains interesting asymptotic behaviors. In
Figure~\ref{fig3}, we illustrate the nature of this function by
fixing the range of the jet opening angle for different value of
$s$. Here we take $\theta_j$ lying between 0.05 and 0.7 radians
($\approx 2.8^\circ$ -- $ 40^\circ$). This figure shows that the
fraction of bursts with small jet opening angles $\theta_j\ll 1$
increases with decreasing values of $s < 0$. Therefore most bursts
have large jet opening angle when $s > 0$. Since beaming effects
are only important for negative $s$, and because beaming effects
are certainly important for GRBs, we only consider negative value
of $s$ in our analysis. Furthermore, from equation~(\ref{eq13}) it
is clear that if $s \leq -1$, it is necessary to have a minimum
opening angle $\theta _j = \arccos\mu_j > 0$ in order to avoid a
divergent fraction of GRBs with small jet opening angles. Note
also that $g(\mu_j)$, eq.~(\ref{eq12}), approaches a
$\delta$-function when $\mu_{\rm jmin}\rightarrow \mu_{\rm jmax}$.

\begin{figure}[htp]
\vskip+0.1in \hskip-0.46in
\includegraphics[width=3.8in]{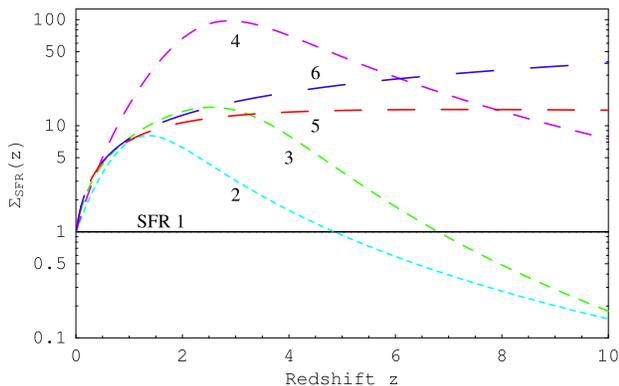}
\vskip-2.2in

\caption{Star formation rate (SFR) function for GRBs, assumed to
be proportional to comoving SFR histories as shown. The solid line
(SFR1) is a constant comoving density; SFR2 and SFR4 are the lower
and upper SFRs given by eq.\ (\ref{eq13}); SFR3 is
the~\citet{hb06} SFR history given by eq.\ (\ref{eq14}), with
($a1=0.015$, $a2=0.1$, $a3=3.4$, $a4=5.5$); and SFR5 ($a1=0.015$,
$a2=0.12$, $a3=3.0$, $a4=1.3$) and SFR6 ($a1=0.011$, $a2=0.12$,
$a3=3.0$, $a4=0.5$) are the SFRs that give a good fit to the
\Swift~and pre-\Swift~redshift distribution.}

\label{fig4} \vskip+0.07in
\end{figure}

It has been suggested  \citep[e.g.,][]{tot97,nat05}
that the GRB formation history is expected to follow
the cosmic SFR ($\Sigma_{_{\rm SFR}}$) derived from the
blue and UV luminosity density of distant galaxies, though with
differences related to the metallicity dependence and fraction of
the subset of high-mass stars that are GRB progenitors. However,
as we show in \S~3, in order to obtain the best fit to
the \Swift~redshift and the pre-\Swift~redshift and jet opening
angle samples, we have to employ a GRB formation history that
displays a monotonic increase in the SFR at high redshifts (SFR5
and SFR6; see Fig.~\ref{fig4}). To investigate the nature of the
GRB formation history, in this work we also assume that
$\dot{n}_{co}(z) \propto \Sigma_{_{\rm SFR}}(z)$, and consider
specific functional forms for $\Sigma_{_{\rm SFR}}(z)$. Due to the
large range of uncertainties in the SFR at $z \gtrsim 1$, we first
consider three different functional forms for the SFR. For the
lower SFR (LSFR) and upper SFR (USFR) that should bound the actual
SFR, we consider the analytic function
\begin{equation}
\Sigma_{_{\rm SFR}}(z) = \frac{1+a_1}{(1+z)^{-a_2} +
a_1(1+z)^{a_3}} \ , \label{eq14}
\end{equation}
\citep{wda04}, with $a_1 = 0.005 (0.0001)$, $a_2 = 3.3(4.0)$, and
$a_3 = 3.0(3.0)$ for the LSFR (USFR).  The LSFR (SFR2) and USFR (SFR4)
describe extreme ranges of optical/UV measurements without and with
dust extinction~\citep{bla99} corrections, respectively.  We also
consider the SFR history (SFR3) of~\citet{hb06}, which is intermediate
between the LSFR and USFR, using their analytic fitting profile given
by
\begin{equation}
\Sigma_{_{\rm SFR}}(z) = \frac{1 + (a_2 z/a_1)}{1+ (z/a_3)^{a_4}}
\ , \label{eq15}
\end{equation}
where $a_1 = 0.015$, $a_2 = 0.10$, $a_3 = 3.4$, and $a_4 = 5.5$
are their best fit parameters\footnote{Adjustments to these values
in later versions of the paper by \citet{hb06} produce negligible
differences in our results.}. Figure~\ref{fig4} shows these SFR
functions, with the function normalized to unity at the present
epoch $z = 0$. The constant comoving density, SFR1, is also
plotted.
Note that SFR2, SFR3, and SFR4 are similar when $z \lesssim 1$,
but differ greatly at higher redshifts. We also consider two
alternative functions shown in Fig.\ \ref{fig4}, SFR 5 and SFR6,
that display a monotonic increase in the SFR at high redshifts
representing positive evolution of the SFR history of the universe
to $z \gtrsim 5$.  Later in the analysis we will show that for
self-consistency we have to utilize SFR5 and SFR6 instead of SFR2,
SFR3, and SFR4, to obtain the best fit to the \Swift~redshift, and
the pre-\Swift~redshift and jet opening angle samples.

\subsection{Redshift, Size, and Jet Opening Angle Distributions}

The directional GRB rate per unit redshift with energy flux $>
\fluxthres$ is thus given by
$$
\frac{d\dot{N}(> \fluxthres)}{d\Omega \ dz }  =  \frac{c g_0}{H_0
(2+s)} \frac{d^2_L(z) \ \dot{n}_{co}(z) }{(1+z)^3 \ \sqrt{\Omega_m
(1+z)^3 + \Omega_\Lambda}} \;
$$
\begin{equation}
\times \; \{ [1-\max(\hat\mu_j,\mu_{\rm jmin})]^{2+s} -
(1-\mu_{\rm jmax})^{2+s}\}\;, \label{eq16}
\end{equation}
after substituting eq.~(\ref{eq12}) into eq.~(\ref{eq11}) and
solving; here
\begin{equation}
\hat\mu_j \; \equiv \; 1 - \frac{\Estarg}{4 \pi d^2_L(z) \ \Delta
t_* \ \fluxthres \lambda_b}\;.
\label{eq17}
\end{equation}
When $f_\e = \fluxthres$, where $\fluxthres$ is the $\nu F_\nu$ flux
threshold sensitivity of a GRB telescope that observes in a waveband
centered at $\e = \bar \e$, eq.~(\ref{eq16}) gives the redshift
distribution. The size distribution of GRBs in terms of their $\nu
F_\nu$ flux $f_\e$ is then simply
$$
\frac{d\dot{N}(> \fluxthres)}{d\Omega  }  =
\frac{c g_0}{H_0
(2+s)} \int_0^{\zmax} dz\;\frac{d^2_L(z) \ \dot{n}_{co}(z)
}{(1+z)^3 \ \sqrt{\Omega_m (1+z)^3 + \Omega_\Lambda}} \;
$$
\begin{equation}
  \times \;\; \{ [1-\max(\hat\mu_j,\mu_{\rm jmin})]^{2+s} - (1-\mu_{\rm jmax})^{2+s}\}\;.
\label{eq18}
\end{equation}
The maximum redshift $\zmax$ can be obtained by solving the
luminosity distance $d_L(z)$ using equation~(\ref{eq17}) to obtain
\begin{eqnarray}
d^2_L(\zmax) & \leq & \frac{\Estarg}{4 \pi (1 - \mu_{\rm jmin}) \
\Delta t_* \ \fluxthres \lambda_b}  \ . \label{eq19}
\end{eqnarray}
Finally, the jet opening angle distribution (see eq.~[\ref{eq11}])
is given by
\begin{eqnarray}
\frac{d\dot{N}(> \fluxthres)}{d\Omega d\mu_j} & = & \frac{c}{H_0}
g(\mu_j)(1-\mu_j) \\
& \times & \; \int_0^{\zmax(\mu_j)}dz\; \frac{d^2_L(z) \
\dot{n}_{co}(z)}{(1+z)^3 \ \sqrt{\Omega_m (1+z)^3 +
\Omega_\Lambda}} \ , \nonumber \label{eq20}
\end{eqnarray}
where $g(\mu_j)$ is given in equations~(\ref{eq12})
and~(\ref{eq13}), and the upper limit of the redshift integration
is determined by equation~(\ref{eq19}).

\section{RESULTS}

In our GRB model, there are seven adjustable parameters: the $\nu
F_\nu$ spectral power-law indices $a$ and $b$, the power-law index $s$
of the jet opening-angle distribution, the range of the jet opening
angles $\theta_{\rm jmin}$ and $\theta_{\rm jmax}$, the absolute
emitted $\gamma$-ray energy $\Estarg$, and the detector threshold $\fluxthres$. As
already mentioned, we consider, for simplicity, a flat $\nu F_\nu$ GRB
SED, so that $a = b = 0$, leaving only the bolometric correction
factor $\lambda_b$, which is set equal to 5.  The remaining parameters
$\theta_{\rm jmin}$, $\theta_{\rm jmax}$, $s$, and $\Estarg$ are
constrained by observations, and the detector properties are used to
estimate $\fluxthres$.

Besides these parameters, we must also assume a GRB comoving rate
density $\dot{n}_{co}(z)= \dot n_{co} \Sigma_{SFR}(z)$, which is
also constrained by fitting our model to the observed data.
Adopting the uniform jet model, we assume that the energy per
solid angle is roughly constant within the well-defined jet
opening angle~\citep[e.g.,][]{fra01,blo03}. With these
assumptions, we investigate the behavior of the redshift and
opening angle distributions. Before we proceed to compare our
model of redshift and jet opening angle distributions with the
samples obtained by the \Swift~and pre-\Swift~instruments, we need
to explore some of the above parameter space.
\begin{figure}[htp]
\vskip+0.in \hskip-0.46in
\includegraphics[width=3.8in]{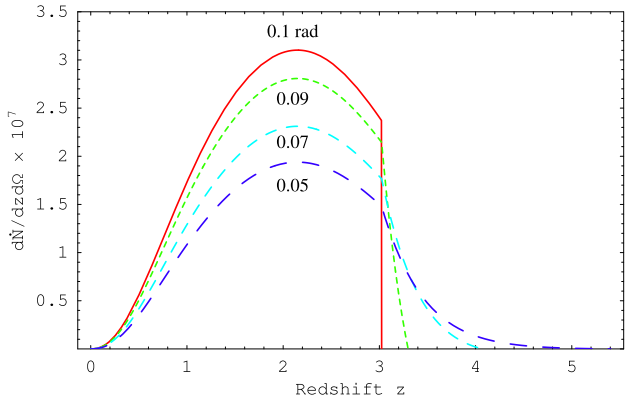}
\vskip-2.2in

\caption{Directional event rate per unit redshift [one event per
$(10^{28} \ \rm cm)^3$ per day per $sr$ per $z$] for an assumed
$\delta$-function distribution of the jet opening angles, so that
$g(\mu_j)=\delta(\mu_j - \hat\mu_j)$, with $\hat \theta_j = 0.1 \
\rm radian$ (solid-curve), using SFR3 and $\fluxthres = 10^{-7} \
\rm ergs \ cm^{-2} \ s^{-1}$. Also shown are the directional event
rates with $\theta_{\rm jmax} = 0.1$ radian, and $\theta_{\rm
jmin} = 0.05$, $0.07$, and $0.09$ radians, with $s = 0$. }

\label{fig5}
\end{figure}

\subsection{Parameter Study of GRB Model}

For illustrative purposes, we utilize SFR3 \citep{hb06} for the
GRB comoving rate density. Furthermore, we also assume that $s =
0$, $\theta_{\rm jmin} = \theta_{\rm jmax} = \theta_j = 0.1$
radian (this assumption implies $g(\mu_j)$ is a $\delta$-function,
see eqs.~[\ref{eq11}], [\ref{eq12}], and Fig.~\ref{fig5}), $\Delta
t_* = 10 \ \rm s$, $\Estarg = 2 \times 10^{51}$ ergs, and
$\fluxthres = 10^{-8}$ and $10^{-7} \ \rm ergs \ cm^{-2} \ s^{-1}$
for the flux sensitivity of detectors for \Swift~and before
\Swift, respectively. We choose $\Delta t_*$ = 10 s, and $\Estarg
= 2 \times 10^{51}$ ergs because these are the mean values of the
GRB duration measured with BATSE assuming that BATSE GRBs are
typically at $z\approx 1$, and the mean beaming corrected
$\gamma$-ray energy release (see Fig.~1), respectively.  We shall
use $\Delta t_* = 10 \ \rm s$ throughout this work. The
calculation is normalized to a local ($z\ll 1$) event rate of one
GRB per day per $(10^{28}{\rm cm})^3$, so that the normalization
of the local rate density is $\dot{n}_{co} = 1.157 \times 10^{-89}
\ {\rm cm^{-3} \ s^{-1}}$.

With the jet opening angle $\theta_j = 0.05$ and $\theta_j = 0.1$
radians, and the detector threshold $\fluxthres = 10^{-7} \ \rm
ergs \ cm^{-2} \ s^{-1}$, the maximum redshifts from which model
GRBs could be detected are $\zmax=5.4$ and $3.02$, respectivley
(see eq.~[\ref{eq19}]). The directional event rate per unit
redshift is shown in Figure~\ref{fig5}. Setting $\theta_{\rm jmax}
= 0.1 \ \rm radian$, and $\theta_{\rm jmin} = 0.05$, $0.07$,
$0.09$ radians (the three dashed curves in Figure~\ref{fig5}), we
plot the directional event rate per unit redshift for different
values of $\theta_{\rm jmin}$ (see eq.~[\ref{eq18}]). As
$\theta_{\rm jmin} \rightarrow \theta_{\rm jmax}$, the directional
event rate per unit redshift of the respective $\theta_{\rm
jmin}$-curves approaches the directional event rate per redshift
of the $\delta$-function $g(\mu_j)$ with $\theta_j = 0.1$ radian.
\begin{figure}[htp]
\vskip+0.05in \hskip-0.46in
\includegraphics[width=3.8in]{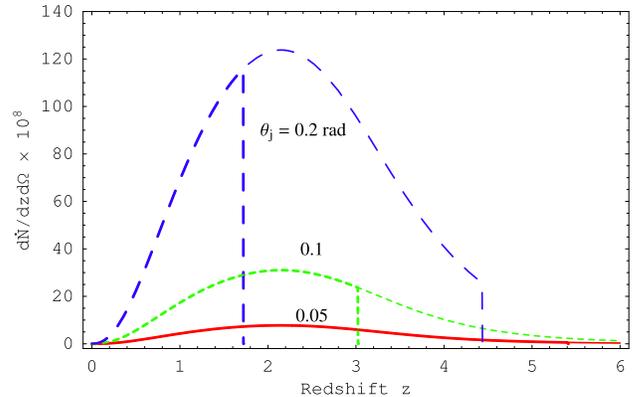}
\vskip-2.2in

\caption{Directional event rate per unit redshift, in units
of GRBs per s per sr per unit redshift $z$, normalized
to one event per
$(10^{28} \ \rm cm)^3$ per day, with $\theta_j =
0.05$, $0.1$, and $0.2 \ \rm radian$, using SFR3. The thin and
thick curves represent the instrument's energy flux sensitivity of
$10^{-8} \ \rm ergs \ cm^{-2} \ s^{-1}$ (\Swift) and $10^{-7} \
\rm ergs \ cm^{-2} \ s^{-1}$ (pre-\Swift), respectively.}

\label{fig6}
\end{figure}
In Figure~\ref{fig6} we examine three different values of the
jet opening angle, $\theta_{\rm jmin} = \theta_{\rm jmax} =
\theta_j = 0.05$, $0.1$, and $0.2$ radians, with two values of
flux sensitivity, namely $10^{-8}$ and $10^{-7} \ \rm ergs \
cm^{-2} \ s^{-1}$. As can be seen, the smaller the jet opening
angle the larger the maximum redshift from which the GRB is
detectable. The maximum observable redshift $z$ associated with a
particular opening angle is indicated by the vertical line in each
curve. Moreover, Figure~\ref{fig6} also shows that whatever the
lower sensitivity telescope detects will also be detected by a
telescope with better sensitivity, because a flat $\nu F_\nu$
spectrum is considered. Thus \Swift, being more sensitive than
earlier telescopes, would in this model detect a sample of GRBs
encompassing those detected with instruments before \Swift.
Spectral behavior can be considered in more detailed treatments
and will change the relations for maximum distances, though
generally with better sensitivity the telescope can view to larger
distances, especially when the more sensitive telescope can
observe lower energy photons.

\begin{figure}[htp]
\vskip+0.1in \hskip-0.46in
\includegraphics[width=3.8in]{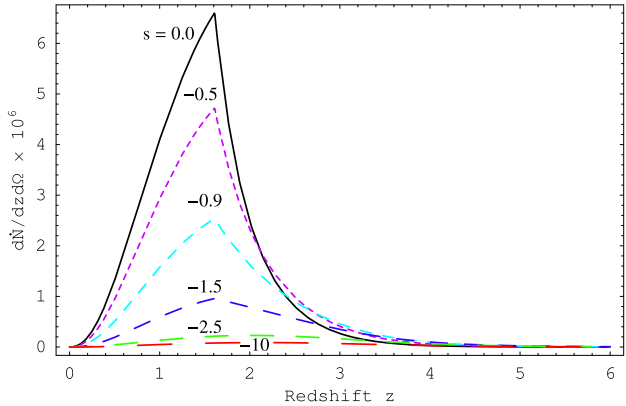}
\vskip-2.2in

\caption{Directional event rate per unit redshift [one event per
$(10^{28} \ \rm cm)^3$ per day per $sr$ per $z$] with $\theta_{\rm
jmin} = 0.05 \ \rm rad$ and $\theta_{\rm jmax} = 0.7 \ \rm rad $,
$s = 0.0$, $-0.5$, $-0.9$, $-1.5$, $-2.5$, and $-10.0$, with SFR3
and $\fluxthres = 10^{-8} \ \rm ergs \ cm^{-2} \ s^{-1}$.}
\label{fig7}
\end{figure}

By varying the opening angle power-law index parameter $s$ over
values $s = 0.0$, $-0.5$,$-0.9$, $-1.5$, $-2.5$, and $-10.0$, with
$\theta_{\rm jmin} = 0.05$, $\theta_{\rm jmax} = 0.7 \ \rm radians
$ and $\fluxthres = 10^{-8} \ \rm ergs \ cm^{-2} \ s^{-1}$, while
keeping all other assumed parameters as before, Figure~\ref{fig7}
shows that the burst rate per unit redshift is progressively
distributed to higher $z$, as expected, since the more negative
$s$ is, the larger is the fractional number of bursts that occurs
at small jet-opening angles $\theta_j$ (see also
Figure~\ref{fig3}). More importantly, as $s \ll 0$, the
directional event rate per unit redshift reaches an asymptotic
limit (see plots for $s=-2.5$ and $-10.0$ in Figure~\ref{fig7}).
\begin{figure}[htp]
\vskip+.10in \hskip-0.46in
\includegraphics[width=3.8in]{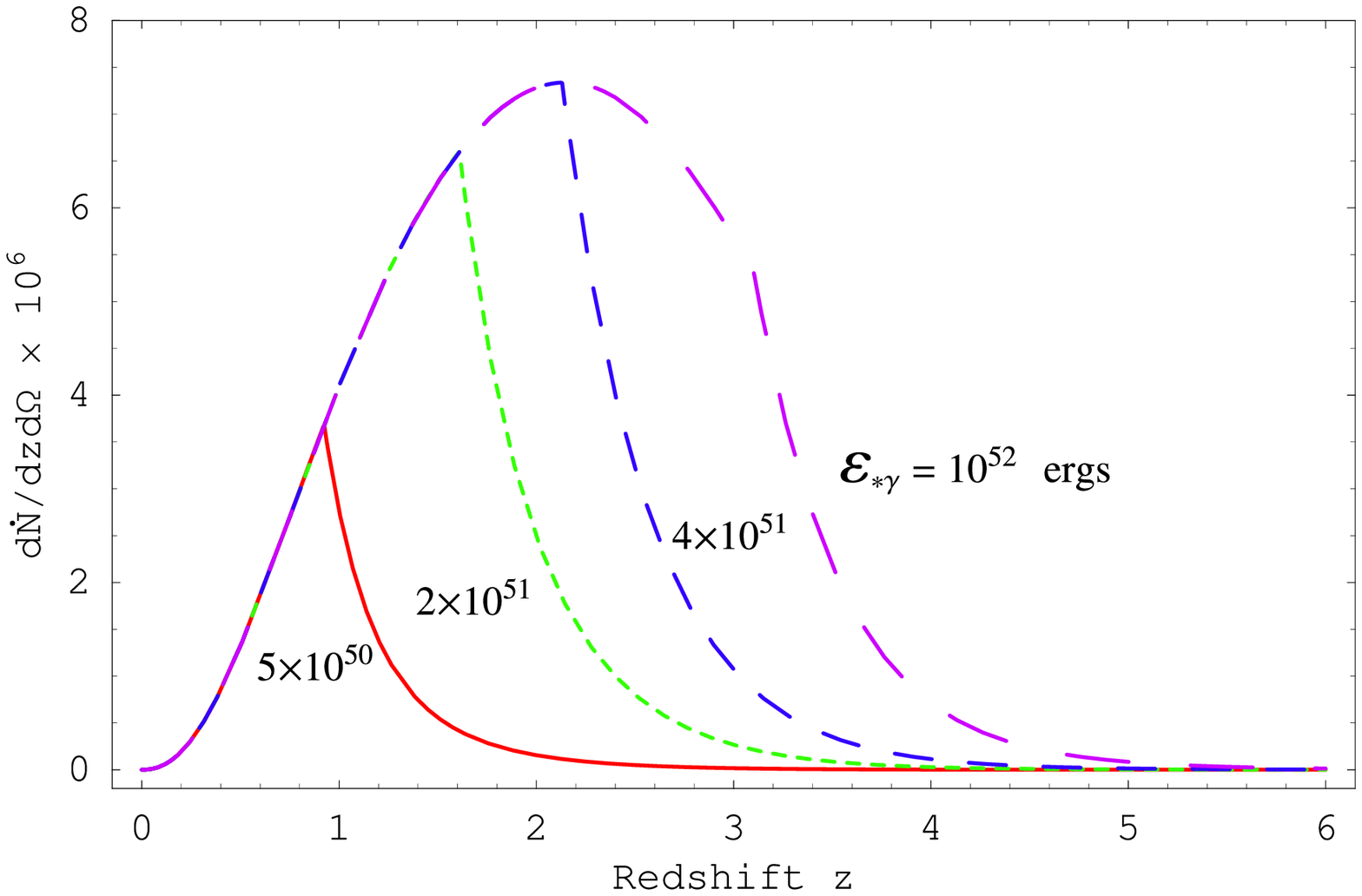}
\vskip-2.2in

\caption{Directional event rate per unit redshift [one event per
$(10^{28} \ \rm cm)^3$ per day per $sr$ per $z$] with $\theta_{\rm
jmin} = 0.05 \ \rm rad$ and $\theta_{\rm jmax} = 0.7 \ \rm rad $
for $\Estarg = 5 \times 10^{50}$, $3 \times 10^{51}$, and $10^{52}
\ \rm ergs$, using SFR3.}
\label{fig8}
\end{figure}

The final parameter that we need to explore is the corrected
$\gamma$-ray energy $\Estarg$. By setting $\Estarg = 5 \times
10^{50}$, $2 \times 10^{51}$, $4 \times 10^{51}$, and $10^{52} \
\rm ergs$ with $\theta_{\rm jmin} = 0.05$ and $\theta_{\rm jmax} =
0.7 \ \rm radians$, $s=0$, and $\fluxthres = 10^{-8} \ \rm ergs \
cm^{-2} \ s^{-1}$, Figure~\ref{fig8} shows that as the emitted
$\gamma$-ray energy increases, the bursting rate is shifted to
higher $z$, as expected.

\subsection{Parameter Fits to the \Swift~and pre-\Swift~\\
Redshift samples}

From~\citet{fb05}, the pre-\Swift~jet opening angle distribution
extends from $\theta_j = 0.05$ to $\theta_j = 0.6$ radians.
However, other workers have reported that the observed jet-opening
angles are as large as $\theta_j = 0.7 \ \rm radians$
\citep[e.g.,][]{gpw05}. Hence, in this work we explore different
ranges of $\theta_j$, ranging from $\theta_{\rm jmin} = 0.05$
radians to $\theta_{\rm jmax} = \pi/2$, $0.7$, and $0.4 \ \rm
radians$. We also use $\Delta t_* = 10 \ \rm s$ as the average GRB
duration in the stationary frame. We first consider SFR3
\citep{hb06} to fit the redshift distribution from the \Swift~and
pre-\Swift~samples, and the opening angle distribution from the
pre-\Swift~sample.  A large number of jet
opening angles associated with the \Swift~redshift sample is
lacking, so our calculated jet opening angle distribution will
be used as a prediction for \Swift. As indicated in
Figure~\ref{fig1}, the corrected $\gamma$-ray energy release is
broadly distributed around the mean value of $\Estarg=2 \times
10^{51} \ \rm ergs$; hence we can explore different values of
$\gamma$-ray energy that best fit the observed redshift and
opening angle samples. It is important, however, that the best-fit
value is not too far off from the mean value in order for our model
to be in accord with observations. Later in the analysis we will
show that our best fit corrected $\gamma$-ray energy $\Estarg$ is
indeed only a factor of 2 larger than the mean.

In Figures~\ref{fig9}, \ref{fig10} and~\ref{fig11}, we plot the
calculated redshift distribution at three different corrected
total $\gamma$-ray energies, $\Estarg = 2 \times 10^{51}$, $5
\times 10^{51}$, and $10^{52} \ \rm ergs$, respectively. In each
panel, each curve represents the calculated cumulative redshift
distribution for the power-law index $s$ of the jet opening angle
distribution.  The values $s=-0.6$, $-1.2$, $-1.8$, and $-2.4$ are
plotted, from far left to far right, respectively.  In each
figure, from top to bottom, we have $\theta_{\rm jmin}=0.05$ and
$\theta_{\rm jmax}=\pi/2$ radians, $\theta_{\rm jmin}=0.05$ and
$\theta_{\rm jmax}=0.7$ radians, and $\theta_{\rm jmin}=0.05$ and
$\theta_{\rm jmax}=0.4$ radians, respectively, as the jet opening
angles. The cumulative redshift distribution (bold solid curves)
of the \Swift~and pre-\Swift~samples are plotted for comparison.
We do not plot $s < -2.4$, since there is little variation in the
redshift distribution (see Fig.~\ref{fig7}). The results in
Figures~\ref{fig9}, \ref{fig10}, and~\ref{fig11} clearly show that
the redshift distributions for different values of $s$ from the
model do not fit the \Swift~redshift sample. Furthermore, the fits
to both the \Swift~and pre-\Swift~samples do not correlate
together with the associated range of the jet opening angle. In
fact it indicates that we need larger burst rates at higher
redshifts ($z > 3$). These results rule out the lower SFR, SFR2,
as can be seen in Figure~\ref{fig12}.

It is important to note at this point that the fit to the
pre-\Swift~and \Swift~redshift distribution should rely on a
single comoving rate density of GRB with the same physical
parameters. If one particular GRB rate model cannot fit the
observed pre-\Swift~and \Swift~redshift distribution at the same
time, then that GRB model has to be eliminated, unless additional
complications are introduced into the model, such as the change of
GRB model parameters with time.

The comparisons also rule out the possibility that the upper SFR
explains the GRB data, even though SFR4 has a higher SFR than SFR2
at large redshifts. In Figure~\ref{fig13} we fit the
pre-\Swift~and \Swift~redshift samples using three different
$\gamma$-ray energies, namely $\Estarg = 5 \times 10^{50}$, $2
\times 10^{51}$, and $10^{52} \ \rm ergs$, with $s=-0.6$, $-1.2$,
$-1.8$, and $-2.4$ using SFR4.  The results for the range of the
jet opening angle $\theta_{\rm jmin}=0.05$, and $\theta_{\rm
jmax}=0.7$ radians show that the calculated redshift distribution
between \Swift~and pre-\Swift~are anticorrelated, that is, smaller
values of $s$, which would improve the fit for \Swift~GRBs, makes
the fit for pre-\Swift~GRBs worse. However, with $s=-0.6$ and
$\Estarg = 10^{52}$ ergs (two bottom panels), the redshift
distributions are marginally acceptable by the model, noting that
the Kolmogorov-Smirnov one-sided test statistic for a sample of 41
and 31 GRBs at the 10\% level are 0.167 and 0.187, respectively
(that is, only 10\% of randomly chosen samples should exhibit
maximum vertical deviations from the model cumulative
distribution; if it is greater than these values, then the model
is an inadequate representation of the data at the 90\% confidence
level). Only 2.5\% of randomly chosen samples should have vertical
excursions greater than 0.212 and 0.238 for a sample of 41 and 31
GRBs, respectively.

When we calculate the opening angle distribution to fit the
pre-\Swift~opening angle sample in Figure~\ref{fig14}, however,
the fit is adequate for $\Estarg = 10^{52}$ ergs and $s = -1.2$
but not for $s = -0.6$.  Hence, the upper SFR (SFR4) is statistically
eliminated.  Nevertheless, the results in
Figure~\ref{fig13} suggest that a greater GRB burst rate at larger
redshifts is required to fit the redshift distributions. The
cumulative jet opening angle distribution (bold solid curves) of
the pre-\Swift~sample is plotted for comparison, but not for
\Swift, since the available opening angles for the associated
\Swift~redshift sample are lacking.

As a result, we consider a GRB comoving source rate density that
exhibits positive evolution above redshift $z>3$. We therefore
modify the Hopkins \& Beacom SFR (SFR3) to obtain the comoving
rate-density that attains (1) a constant comoving rate-density at
high redshift $z$ (SFR5), and (2) a monotonically increasing
comoving rate-density with increasing redshift (SFR6) in order to
provide a high GRB burst rate at large $z$ (see caption to
Figure~\ref{fig4} for coefficients of the SFR models).

In Figures~\ref{fig15} and \ref{fig16} we begin to see some
improvement in fitting the pre-\Swift~and \Swift~redshift samples
when using SFR5 and SFR6 with $\Estarg=2 \times 10^{51} \ \rm
ergs$. Nevertheless, the fitting is still somewhat inconsistent
between the pre-\Swift~redshift distribution and the jet opening
angle. For example, in Figure~\ref{fig15} using SFR5, the fitting
to the pre-\Swift~and \Swift~redshift distribution shows a good
fit with $s=-1.2$ for a range of jet opening angles between
$\theta_{\rm jmin}=0.05$  and $\theta_{\rm jmax}=0.4$ radians (two
bottom panels). However, the result from the fitting between the
pre-\Swift~redshift distribution and the opening angle distribution with
$s=-1.2$ is inconsistent (see Figures~\ref{fig15}
and~\ref{fig16}). Above redshift $z>2$, the calculated redshift
distribution underestimates the fractional number of bursts that
should occur at high redshifts $z$. On the other hand, the
calculated opening angle distribution overestimates the fractional
number of GRBs that should occur at small opening angles. Recall
that if a burst occurs at a small opening angle, then this
translates into the possibility of observing this particular burst
at a high redshift $z$. Hence, the fit with the range of the jet
opening angle $\theta_{\rm jmin}=0.05$, and $\theta_{\rm
jmax}=0.4$ radians is not yet acceptable.

In Figures~\ref{fig17} and \ref{fig18} we see similar results when
using SFR6. However, it is interesting to note that we can resolve
this problem if we could increase the fractional number of GRBs at
both high and low redshifts $z$. As a result, this will provide a
self-consistent fitting between the pre-\Swift~redshift and the
jet opening angle distributions, as well as the \Swift~redshift
distribution. This problem can be solved by adjusting the
$\gamma$-ray energy $\Estarg$. Using $\Estarg=4 \times 10^{51} \
\rm ergs$, Figures~\ref{fig19} and \ref{fig20}, and
Figures~\ref{fig21} and \ref{fig22} show a statistically
acceptable fit to the pre-\Swift~and \Swift~redshift distribution
and the pre-\Swift~opening angle distribution with $s=-1.2$, and
the range of the jet opening angle $\theta_{\rm jmin}=0.05$ and
$\theta_{\rm jmax}=0.7$ radians for SFR5 and SFR6, respectively.
For SFR5, the fitting can be improved with $s=-1.3$, as indicated
by the solid curves (see Figures~\ref{fig19} and \ref{fig20}).

The above analyses show that $\Estarg = 4 \times 10^{51} \ \rm
ergs$ gives the best fit to the pre-\Swift~redshift and the jet
opening angle samples, and the \Swift~redshift sample. This
best-fitted $\gamma$-ray energy is interesting since~\citet{fra01}
and~\citet{blo03} show that the standard-energy reservoir is
$\Estarg \cong  10^{51} \ \rm ergs$. However, with a larger
sample set, the data from~\citet{fb05} shows that $\Estarg$ is
broadly distributed between $\Estarg \sim 10^{51-52} \ \rm ergs$,
as shown in Figure~\ref{fig1}. Because we have been able to obtain
an acceptable fit to the data with discrete values of $\Estarg$
and $\Delta t_*$, a more complicated model where a range of values
of  $\Estarg$ and $\Delta t_*$ were allowed would certainly permit
excellent fits to the data.

\subsection{Estimating the Beaming Factor and \\ the Luminosity Function}

We can utilize our best-fitted parameters to estimate the average
beaming factor $<f_b^{-1}>$.  The beaming factor average over the
opening angle distribution is given by
\begin{eqnarray}
<f_{_b}> & = & \frac{\int^{\mu_{\rm jmax}}_{\mu_{\rm jmin}}
(1-\mu_j) g(\mu_j) d\mu_j}{\int^{\mu_{\rm jmax}}_{\mu_{\rm jmin}}
g(\mu_j) \ d\mu_j} \\
& = & \left(\frac{1+s}{2+s}\right)\left[ \frac{(1-\mu_{\rm
jmin})^{2+s} - (1-\mu_{\rm jmax})^{2+s}}{1-\mu_{\rm jmin})^{1+s} -
(1-\mu_{\rm jmax})^{1+s}} \right] \ , \nonumber \label{eq21}
\end{eqnarray}
\vspace{+1mm}
where $g(\mu_j)$ is given in equation~(\ref{eq12}).
Using the best-fitted parameters $s=-1.2$ and $s=-1.3$ and the
range of the jet opening angles between $\theta_{\rm jmin}=0.05$
and $\theta_{\rm jmax}=0.7$, we obtain $<f_b>^{-1}\simeq 34-42$ as
the beaming factor for $s=-1.2$ and $s=-1.3$, respectively. Our
value of $<f_b>^{-1}$ is about a factor of 2 smaller than the
value obtained by~\citet{gpw05}.

We can also infer the luminosity function from the jet opening angle
distribution. From equations (\ref{eq8}) and (\ref{eq9}),
the jet opening angle distribution is $d\dot{N}/d\mu_j =
g(\mu_j)$, which is related to luminosity by the expression
\begin{equation}
\frac{d\dot{N}}{dL_*} = g(\mu_j) \left| \frac{d\mu_j}{dL_*}
\right| , \label{eq23}
\end{equation}
where $L_*$ is the GRB apparent isotropic luminosity, and $g(\mu_j)$ is the
jet opening angle distribution given in equation (\ref{eq12}).
Since the energy flux is defined as $\Phi_E = L_*/4 \pi d^2_L$ or
$\fluxeps = \nu F_\nu = \nu_* L_{\nu *}/4 \pi d^2_L$, then the
isotropic luminosity in the bursting frame is $L_* = \nu_* L_{\nu
*}$, so that $L_* = 4 \pi d^2_L \fluxeps$. Utilizing equation
(\ref{eq17}), the jet opening angle in terms of isotropic
luminosity is given by
\begin{equation}
\mu_j = 1 - \frac{\Estarg}{\Delta t_* \lambda_b L_*} \ .
\label{eq24}
\end{equation}
Substituting equation (\ref{eq24}) into equation (\ref{eq23}), we
obtain $dN/dL_* \propto L^{-(s+2)}_*$ after utilizing the jet
opening angle distribution in equation (\ref{eq12}). Since our
best fit to the pre-\Swift~redshift and opening angle samples and
the \Swift~redshift sample constrains the jet opening angle
distribution power-law index $s$ between $-1.3$ and $-1.2$ for
using SFR5 and SFR6, respectively, our apparent isotropic
luminosity function is $\propto L^{-3.25}_*$ within the uniform
jet model. In the universal structured jet model, the luminosity
function is suggested to be $\propto L^{-2}_*$
\citep[e.g.,][]{gpw05,psf03}; others have suggested that the
luminosity function is $\propto L^{-2.3}_*$
\citep[e.g.,][]{lfr02}. For our best fitting model with jet
opening angles ranging from $\theta_j = 0.05$ to $\theta_j = 0.7$
rad, this implies bolometric $\gamma$-ray luminosities in the
range $\approx 2\times 10^{51}$ to $\approx 3\times 10^{53}$ ergs
s$^{-1}$.

The results for our best fitting model\ are shown in
Figure~\ref{fig23}. The fraction of high-redshift GRBs
in the high-redshift universe implied
by our fits using SFR5 and SFR6 is
estimated at 8 -- 12\% and 2.5 -- 6\% at $z \geq 5$
 and $z\geq 7$, respectively. If this fraction is not observed with
Swift, then it suggests that SFR5 and SFR6 do not continue
with positive evolution $z \gg 5$. Indeed, it is likely that
the formation rate of GRBs at high redshift declines at
sufficiently high redshift.
We can also derive the fractional number of GRBs with $z
\leq 0.25$ from Figure~\ref{fig23}. We find that  $\lesssim 0.2$\% of
long duration GRBs should occur in the low-redshift ($z \leq 0.25$)
universe.

\section{DISCUSSION AND CONCLUSIONS}

In this work, we considered whether the differences between the
pre-\Swift~and \Swift~redshift distributions can be explained with
a physical model for GRBs that takes into account the different
flux thresholds of GRB detectors.  The model presented here
parameterizes the jet opening angle distribution for an assumed
flat $\nu F_\nu$ spectrum,
and finds best fit values for the $\gamma$-ray energy release for
different functional forms of the comoving rate density of GRBs, assuming
that the properties of GRBs do not change with time. Adopting
the uniform jet model, we assumed that the energy per solid angle
is roughly constant within a well-defined jet opening angle. The
pre-\Swift~redshift sample suggests that the range of the jet
opening angle is between $0.05$ and $0.7$ radians. We explored
three different ranges of jet opening angle with $\theta_{\rm
jmin}=0.05$, and $\theta_{\rm jmax}=\pi/2$, $0.7$, and $0.4$
radians.

The results in \S~3 show that an intrinsic distribution in the
jet opening angles and single values for all other parameters
yields a good fit to the pre-\Swift~and
\Swift~redshift samples, and furthermore provides an acceptable fit to the
distribution of opening angles measured with
pre-\Swift~GRB detectors. A good fit was only possible, however,
by modifying the Hopkins \& Beacom SFR to provide positive evolution of the SFR
history of GRBs to high redshifts (SFR5 and SFR6; see Fig.~\ref{fig4}).
The best fit values were
obtained with $\Estarg = 4\times 10^{51}$ ergs, $\theta_{\rm
jmin}=0.05$, $\theta_{\rm jmax}=0.7$ radians, and $s \cong
-1.25$. The best-fit value for the absolute $\gamma$-ray energy released by
a GRB is about a factor of 2 larger than the mean value of the \citet{fb05} sample of
pre-\Swift~GRBs.

The results of our fitting therefore indicate that GRB activity
was greater in the past and is not simply proportional to the bulk
of the star formation as traced by the blue and UV luminosity
density of the universe. This is contrary to the conclusion of
\citet{pm01} based upon a study of BATSE size distribution and a
small number of GRBs with known redshifts in the pre-Swift era.
They obtain worse fits for SFRs that increase at high redshift.
The BATSE analysis of \citet{mbbh06} gives results that are
consistent with the required monotonically increasing GRB SFR that
we find. \citet{drm06} also conclude that a rising GRB SFR is
implied, and also note that the star formation rate cannot
increase in proportion to greater past GRB activity, as this would
result in overproduction of metals. Changes in the progenitor
population or GRB properties, for example, due to changing
metallicities with time, might account for the large number of
high-redshift GRBs observed with Swift, but this would introduce a
large number of unconstrained parameters. Moreover, one would then
have to explain why the Amati and Ghirlanda correlations relating
spectral properties and energy releases are insensitive to
redshift.

Assuming that the  GRB properties do not change with time, we
therefore find that the GRB source rate density must display
positive evolution to at least $z \gtrsim 5$. Why would GRB
activity be greater in the past? One possibility suggested by
recent studies \citep{sta06,fru06} on the metallicities of host
galaxies of GRBs is that GRBs derive from progenitor stars with a
lower metallicity than the progenitors of Type II SNe. In this
case, one might expect that the peak of GRB activity would occur
earlier in the universe than during the peak of star formation
activity at $z \approx 2$ -- 3. Nevertheless, the
high-redshift fraction of GRBs detected with Swift will test the
validity of SFR5 and SFR6 used to model the GRB data.

The models that fit the data imply that 8 -- 12\% of Swift GRBs
occur at $z \gtrsim 5$, in accord with the data as shown in
Figure~\ref{fig23}. Our model predicts that 2.5 -- 6\% of GRBs
should be detected from $z\geq 7$. If no
 GRB with $z > 7$ is detected by the time $\approx 100$ Swift GRBs
with measured redshifts are found, then we may conclude that the
GRB formation rate has begun to decline above $z\approx 5$. By
contrast, if a few $z\gtrsim 7$ GRBs have been detected within the
next year or two with Swift, then this would be in accord with our
model and would support the conjecture of \citet{bl06} that a
second episode of Population I and II star formation took place at
$z \gtrsim 5$, possibly even including some GRBs that originate
from Population III stars. Our model parameters suggest that GRBs
can be detected with Swift to a maximum redshift $z \approx 20$,
even given the limits placed on the range of jet angles, in
particular, $\theta_{min} \geq 0.05$.

The relationship between the galaxies that host GRBs and their
metallicities remains controversial, with evidence for
high-redshift, low-metallicity GRB host galaxies \citep{fru06}
countered by examples of GRBs found in galaxies with moderate
metallicity \citep{ber06,fyn06a}. \citet{sta06} find that 5 GRBs,
or several per cent of the GRB population with known redshifts,
are found at low ($z\lesssim 0.25$) redshift in galaxies that have
very low metallicity compared with Solar metallicity and with the
distribution of metallicities of low-redshift galaxies in the
Sloan Digital Sky Survey. In 4 out of the 5 low-$z$ GRBs,  the
apparent isotropic energy release of these GRBs is much smaller
than $10^{51}$ ergs, and so these GRBs are unrepresentative of
typical long duration GRBs. Moreover, the large number of these
low redshift GRBs already means that they are not typical of the
sample of GRBs considered in this paper, as the model prediction
is that $\approx 0.14$ -- 0.2\%   of the ``classical" GRBs should
be detected at low redshifts.\footnote{A value of $\approx 0.3$\%
can be estimated by considering the fractional volume within $z =
0.25$ compared with $z\cong 1$, and that the SFR activity at $z
\cong 1$ was a factor $\cong 5$ greater than at $z<0.25$.} These
low-redshift GRBs may also belong to a completely distinct
population of GRBs compared to their high-redshift counterparts,
as also suggested by the differing luminosity functions of the two
populations \citep{lia06}. Alternatively, the low-redshift GRBs
could be normal GRBs that appear weak because, for example, their
jetted emission is observed off-axis \citep[e.g.][]{yam03}.

Although we have restricted our treatment to the long-duration,
soft-spectrum class of GRBs, the separation between the short hard
and long soft class of GRBs is not distinct. Moreover, recent
observations of nearby GRBs show that the nearby ($z = 0.125$) GRB
060614, a long duration GRB in terms of its light curve
\citep{geh06}, lacks a supernova excess \citep{fyn06,val06}, does
not fit neatly into either the short hard or long soft classes
\citep{zha06}, and could represent a separate population of GRBs
\citep{gal06}. Phenomenological studies have also indicated that
the long GRB population is bimodally distributed in terms of the
autocorrelation function of the light curves
\citep[e.g.,][]{bor04}. \citet{tav98} argues for multiple
populations by examining the size distributions of different
groups of BATSE GRBS separated according to spectral properties.
From an analysis of durations of BATSE, \citet{hor06} identifies
an intermediate duration population that displays the softest
spectra and comprises $\approx 11$\% of the total BATSE GRBs.

That we were able to obtain good fits to the different data sets
using a single GRB model may indicate that the contribution of
sub-populations to the total number of GRBs is small. Alternately,
the model parameters might have been adjusted to take into account
the aggregate long duration GRBs. This can be tested in future
work using a larger Swift data set that allows subgroups to be
separately studied based on spectra or durations.

Figure~\ref{fig24}a shows the model integral size distribution
(see eq.~[\ref{eq18}]) of GRBs predicted by our best fit model.
The plots are normalized to the current total number of observed
GRBs per year from BATSE, which is 550 bursts per $4\pi$ sr
exceeding a peak flux of 0.3 photons $\rm cm^{-2} \ s^{-1}$  in
the $\Delta E=50-300$ keV band for the $\Delta t=1.024$ s trigger
time \citep{ban02}. From the model size distribution, we find that
$\cong 340$ to 360 GRBs per year should be detected with a
BATSE-type detector over the full sky above an energy flux
threshold of $\sim 10^{-7} \ \rm ergs \ cm^{-2} \ s^{-1}$, or
photon number threshold of $0.625 \ \rm ph \ cm^{-2} \ s^{-1}$.
This range of values is determined by the two model fits with SFR6
and SFR5, respectively.  We estimate from our fits that $\approx
1190$ --  1370 GRBs take place per year per $4 \pi$ sr with a flux
$\gtrsim 10^{-8} \ \rm ergs \ cm^{-2} \ s^{-1}$, or $0.0625 \ \rm
ph \ cm^{-2} \ s^{-1}$. The field of view of the BAT instrument on
Swift is 1.4 sr \citep{geh04}, implying that Swift should detect
$\approx 130$ -- 150 GRBs per year; currently, \Swift~observes
about 100 GRBs per year. This minor discrepancy may be a
consequence of using a flat $\nu F_\nu$ GRB spectrum in our model.

We also show the differential size distribution of BATSE GRBs from
the Fourth BATSE catalog \citep{pac99} in comparison with our
model prediction in Figure~\ref{fig24}b. As can be seen, our model
gives a good representation of the size distribution of the BATSE
GRB distribution within the statistical error bars, except for a
slight overprediction of the number of the brightest GRBs
\citep[see also][]{drm06,bd00}. These brightest GRBs typically
originate from $z \lesssim 1$, possibly suggesting a slight
reduction in the number of GRBs with $z\ll 1$ compared to the GRB
rate densities considered. Below a photon number threshold of $0.3
\ \rm ph \ cm^{-2} \ s^{-1}$ in the 50 -- 300 keV band, the
observed number of GRBs falls rapidly due to the sharp decline in
the BATSE trigger efficiency at these photon fluxes
\citep[see][]{pac99}. The size distribution of the \Swift~GRBs
will extend to much lower values, $\approx 0.0625 \ \rm ph \
cm^{-2} s^{-1}$, and we can use our model to predict the peak flux
size distribution to fit the \Swift~data, noting however that the
\Swift~triggering criteria are more complicated than a simple rate
trigger, particularly near threshold \citep{ban06}.

We can  derive the burst rate $\dot{\rho}$ in our local universe
by normalizing our best fit models to the BATSE results of 340 --
360 GRBs per year full sky with peak fluxes exceeding $10^{-7}$
ergs cm$^{-2}$ s$^{-1}$. Comparing this value with the rate
calculated from eq.\ (\ref{eq18}) for our best fit models, we
obtain local GRB rate densities of $\dot n_{\rm GRB} = 9.6$
Gpc$^{-3}$ yr$^{-1}$ for SFR5 and $\dot n_{\rm GRB} = 7.5$
Gpc$^{-3}$ yr$^{-1}$ for SFR6. Since the mean volume occupied by
an $L^*$ galaxy in the local universe is $\approx 200$ Mpc$^{-3}$
\citep{lov92,wij98}, we obtain the local event rate in our Galaxy,
assumed to be a typical $L^*$ galaxy, to be $\dot{\rho} \approx
1.9 \times 10^{-6}$ and $1.5 \times 10^{-6}$ events per year per
$L_*$ galaxy, for SFR5 and SFR6, respectively. This implies that about 1
event occurs every 0.6 Myrs in our Galaxy. Comparing this to
the rate of Type Ib/c supernovae of $\approx 1$ every 360 years in
our Galaxy \citep{cap99}, we estimate that the rate of GRBs is
only $\approx 0.06$\% of the rate of SN Ib/c supernovae in the
Galaxy. \citet{ber03} find that $\leq 3$\% of GRBs could originate
from SN Ib/c by comparing the radio brightnesses of SN Ib/c with
that of SN 1998bw associated with GRB 980425. Our results fall
comfortably within the observational limits. We also find that
about 1 out of every 20,000 SNe make a GRB, compared to the value
of 1 out of every $10^5$ -- $10^6$ found by \citet{drm06}.

If GRBs do occur at the rate of more than once per Myr in the
Milky Way, depending on uncertain metallicity effects
\citep{sta06}, then the Earth will intercept the beam of a GRB
about once every 25 Myrs, given the beaming factor $\langle
f^{-1}_b \rangle \sim 40$ obtained in our analysis. If a Milky Way
GRB is located at an average distance of 10 kpc, then the average
fluence deposited by the GRB is on the order of $\approx 10^7$
ergs cm$^{-2}$. Strong astrobiological effects begin to be
important when the GRB fluence is $\approx 10^8$
 -- $10^9$ ergs cm$^{-2}$, or when the GRB occurs within $\approx 1$ -- 2
kpc with its beam pointing in our direction \citep{dh05}. This will
happen perhaps a few times per Gyr, so that GRBs could have significant
effects on terrestrial evolution \citep{mel04,tho05}.

The number of jet opening angles measured with follow-up
observations of \Swift~GRBs is not yet sufficient to produce a
statistically reliable sample. Thus our jet opening angle
distribution shown in Figures~\ref{fig20} and \ref{fig22} with
$s=-1.3$ and $-1.2$, for SFR5 and SFR6, respectively, provide
a prediction for \Swift.  With \Swift~and pre-\Swift~thresholds
of $\approx 10^{-8}$ and $\approx 10^{-7} \ \rm ergs \ cm^{-2} \
s^{-1}$, our model predicts the mean measured jet opening angle $
\langle\theta_j\rangle \approx 0.16$, and $0.12$ rad,
respectively, which can be compared with the measured average
value of $\langle \theta_j\rangle \approx 0.125$ for the
pre-\Swift~sample. Thus we expect to detect more faint
low-redshift, large opening angle GRBs that pre-\Swift~satellites
could not detect \citep[see also][]{psf03}. It is interesting to
note that with a better detector sensitivity, the mean jet angle
is shifted to a higher angle and not vice versa, as might be
expected. This suggests that on average we expect to
see more long-duration bursts with larger opening angle ($\theta_j
> 0.12$ radian), meaning that the mean times for the achromatic
breaks in the light curves could be longer than for the
pre-\Swift~GRBs when one also takes into account the larger mean
redshift of the \Swift~sample.

In conclusion, we have developed a physical model for GRBs and
obtained parameters to the model by fitting the GRB redshift and
opening angle distributions. We could only obtain statistically
acceptable joint fits if the comoving rate density of GRBs
increases monotonically to $z \gtrsim 5$. Our fitting results give
a rate of $\approx 1$ GRB every 600,000 yrs in our Galaxy if
metallicity effects do not play a large role in the galaxies in
which GRBs are formed. In this case, the increase in the GRB rate
density to high redshifts would have a different cause, for
example, an episode of enhanced star formation at $z\gtrsim 5$
\citep{bl06}. The opening angle distribution of GRBs measured with
\Swift, and the fraction of low-$z$ and high-$z$ standard
long-duration GRBs will test this model.

\acknowledgements We thank D.\ Band, J.\ Beacom, E.\ Berger, J.\ Bloom,
and A.\ Friedman for discussions and correspondence, and the referee
for comments, corrections, and useful suggestions. T.~L. is funded
through NASA {\it GLAST} Science Investigation No.~DPR-S-1563-Y.
The work of C.~D.~D. is supported by the Office of Naval Research.

%
%

\appendix

\section{Bursting Rate of Beamed Cosmic Sources in a Flat $\Lambda$CDM Universe}

The differential number of bursts per unit proper time per unit
volume with beaming-corrected absolute
$\gamma$-ray energy release between $\cal{E}_{*\gamma}$ and
$\cal{E}_{*\gamma}$ + $d\cal{E}_{*\gamma}$, jet opening cosine angle
$\mu_j = \cos\theta_j$ between $\mu_j$ and $\mu_j + d\mu_j$,
peak energy between $\epsilon_{pk*}$ and $\epsilon_{pk*} +
d\epsilon_{pk*}$, burst duration between $\triangle
t_*$ and  $\triangle t_* + d(\triangle t_*)$, that occur in volume element
between $V_*$ and $V_* + dV_*$ within observer angle $\mu =
\cos\theta$ between $\mu$ and $\mu + d\mu$ during
time interval between $t_*$ and $t_* + dt_*$ is given by
\begin{equation}
dN = \dot{n}_*(\alpha, t_*) d\alpha d\mu dV_* dt_* \ . \label{A1}
\end{equation}
Here the parameter $\alpha$ represents the various parameters of
the problem, namely $\cal{E}_{*\gamma}$, $\mu_j$,
$\epsilon_{pk*}$, and $\triangle t_*$. Hence, the directional
burst rate per unit observer time $t$ is given by
\begin{equation}
\frac{d\dot{N}}{d\Omega} = \frac{\dot{n}_*(\alpha_, t_*) d\alpha
d\mu dV_*}{d\Omega}\left|\frac{dt_*}{dt}\right| \ . \label{A2}
\end{equation}
In a flat $\Lambda$CDM universe, the Robertson-Walker metric
describes the line element $ds$ for a homogeneous, isotropic
expanding universe, and is given by
\begin{equation}
ds^2 = c^2 dt^2 - R^2(t)\left[\frac{dx^2}{1-k x^2} + x^2
\left(d\theta^2 + \sin^2\theta \ d\phi^2\right)\right] \ ,
\label{A3}
\end{equation}
where $R(t)$ is the dimensionless scale factor of the universe and
($x, \theta, \phi$) are the \emph{comoving coordinates} of a point
at rest in the Hubble flow. The curvature index $k= -1/{\cal
R}^2$, $0$, and $+1/{\cal R}^2$ represents an open, flat, and
closed universe model, respectively, where ${\cal R}$ is the
radius of curvature at the present epoch. At the present epoch, we
let $R(t) = 1$ and $x$ correspond to the physical distance between
two comoving points. The ratio of the scale factor, $R(t_*)$, at
the emitted time, $t_*$, to the scale factor $R = R(t_0)$, at the
present time, $t_0$, is given by,
$$\frac{R_*}{R} = \frac{1}{1+z} = \frac{\lambda_*}{\lambda} \; .$$
The following relations hold from the above definition,
namely
\begin{equation}
1+z =\;{\nu_*\over \nu} = {\Delta t\over \Delta t_*} =
{\epsilon_*\over \epsilon}\;, \label{A4}
\end{equation}
where the last expression applies to photons and relativistic
particles in an expanding universe with $R_* = R(t_*)$, and $R =
R(t_0)$.

To calculate the differential volume element $dV_*$ in equation
(\ref{A2}), we utilize the fact that the product of differential
length elements corresponding to infinitesimal changes in the
coordinates $x$, $\theta$, and $\phi$ equals the differential
volume element $dV$. Hence, the proper volume element of a slice
of the universe at time $t_*$ is, from eq.\ (\ref{A3}),
\begin{equation}
dV_* = \frac{(R_* x)^2 R_* dx d\Omega_*}{\sqrt{1-k x^2}}\;\;
\stackrel{k \rightarrow 0}{\rightarrow} \;\; R_*^3 x^2 dx
d\Omega_*\;, \label{A5}
\end{equation}
specializing to a flat universe, where $d\Omega_*$ represents the
solid angle at the emitted time. Substituting equation (\ref{A5})
into equation (\ref{A2}) for $dV_*$, we obtain
\begin{equation}
\frac{d\dot{N}}{d\Omega} = \dot{n}_*(\alpha, t_*) R_*^3 x^2 dx
d\alpha d\mu \left|\frac{dt_*}{dt}\right| \ , \label{A6}
\end{equation}
where in a homogeneous isotropic expanding
universe with no intervening gravitating matter to
distort the paths of the light rays, $d\Omega_* = d\Omega$.

Light travels along a null
geodesic, $ds=0$, which is a path along which $\theta$ and $\phi$
are constants. Thus in the Robertson-Walker metric the proper time
element elapsed for a photon to propagate a differential comoving
distance at earlier time $t_*$ is just $cdt_* = R_* dx$ or $dx
= c dt_*/R_*$. Upon substituting $dx$ into equation (\ref{A6})
we obtain,
\begin{equation}
\frac{d\dot{N}}{d\Omega} = \frac{c \ \dot{n}_*(\alpha, z) (R_*
x)^2 d\alpha d\mu}{1+z} \left|\frac{dt_*}{dz}\right| dz\ ,
\label{A7}
\end{equation}
where we have also expressed everything in terms of redshift z by
transforming from the variable $t_*$ to $z$.

An expression for $(R_* x)^2$ is derived by recalling the
relationship between energy flux $\Phi_E$ and luminosity distance
$d_L$, namely,
\begin{eqnarray}
{d{\cal E}\over dA dt} = \Phi_E = {L_* \over 4\pi d_L^2} = {d{\cal
E}_*\over 4\pi d_L^2d t_*} \; & = & (4\pi
d_L^2)^{-1}\;\left({d{\cal E_*}\over d{\cal
E}}\right)\;\left({dt\over dt_*}\right)\; \left({d{\cal E}\over
dA dt}\right)\;dA  \nonumber \\
& = & {(1+z)^2 \over 4\pi d_L^2}\;\Phi_E\;dA\; , \nonumber
\end{eqnarray}
or
\begin{equation}
{(1+z)^2 \over 4\pi d_L^2}\;dA = 1 \; , \label{A8}
\end{equation}
where the area element $dA$ of the Robertson-Walker metric is
\begin{eqnarray}
dA = (R x)^2  \sin\theta \ d\theta  d\phi = R^2  x^2 d\Omega \;.
\nonumber
\end{eqnarray}
Hence
\begin{equation}
{dA_*\over dA} = \left({R_* x\over R x}\right)^2
\frac{d\Omega_*}{d\Omega}= {1\over (1+z)^2}\;, \label{A9}
\end{equation}
where we have utilized equation (\ref{A4}) and use $d\Omega_*=d\Omega$.
Substituting
equation (\ref{A9}) into equation (\ref{A8}) for $dA$, we obtain
\begin{equation}
(R_* x)^2 = \frac{d_L^2}{(1+z)^4} \; . \label{A10}
\end{equation}

For a flat $\Lambda$CDM universe \citep{pee93},
\begin{equation}
\left|{dz\over dt_*}\right| = H_0 (1+z) \sqrt{\Omega_m(1+z)^3 +
\Omega_\Lambda}\; , \label{A11}
\end{equation}
where $\Omega_m$ and $\Omega_\Lambda$ are the ratios of the energy
densities of total mass, dominated by dark matter, and dark
energy, respectively, compared to the critical density.
Substituting equations \ (\ref{A10}) and (\ref{A11}) into equation
(\ref{A7}) gives
\begin{equation}
\frac{d\dot{N}}{d\Omega} = \frac{c}{H_0} \frac{d_L^2(z)\;
\dot{n}_*(\alpha,z)\; d\alpha\; d\mu\; dz}{(1+z)^6
\sqrt{\Omega_m(1+z)^3 + \Omega_\Lambda}} \ , \label{A12}
\end{equation}
for randomly oriented GRB sources. After using the relation that
the proper density $\dot{n}_*(\alpha, z)$ increases with redshift
$\propto (1+z)^3$ in the comoving density $\dot{n}_{co}(\alpha,
z)$, that is $\dot{n}_*(\alpha, z) = (1+z)^3\;\dot{n}_{co}(\alpha,
z)$, we obtain the result given in equation (\ref{eq8}).

\clearpage

\begin{figure}[t]
\vskip+.20in \hskip-0.85 in
\includegraphics[width=8in]{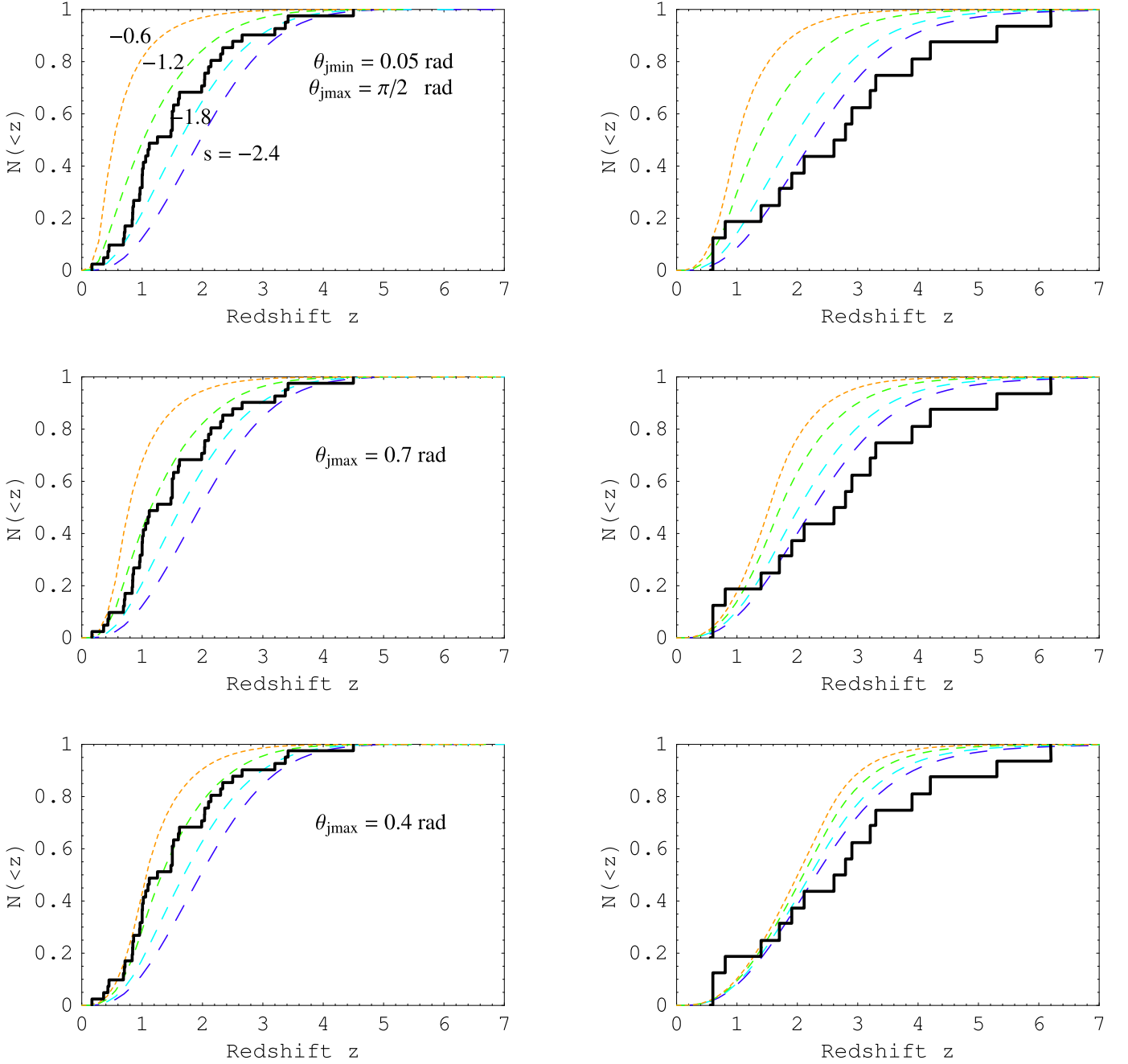}
\vskip-1.8 in

\caption {Measured and model redshift distributions for SFR3
\citep{hb06}. In the left column, from top to bottom, we have the
pre-\Swift~cumulative redshift distribution with the assumed range
of jet opening angles $\theta_{\rm jmin}=0.05$, $\theta_{\rm
jmax}= \pi/2$, $0.7$, and $0.4 \ \rm radians$, respectively. The
right column has results as in the left column but for
\Swift~data. In each panel, each curve represents the cumulative
redshift distribution for different jet opening angle power-law
indices $s$, where $s=-0.6$, $-1.2$, $-1.8$, and $-2.4$, from far
left to far right, respectively. The assumed gamma-ray energy is
$\Estarg = 2 \times 10^{51} \ \rm ergs$.}

\label{fig9}
\end{figure}
\clearpage
\begin{figure}[t]
\vskip+.2in \hskip-0.85in
\includegraphics[width=8in]{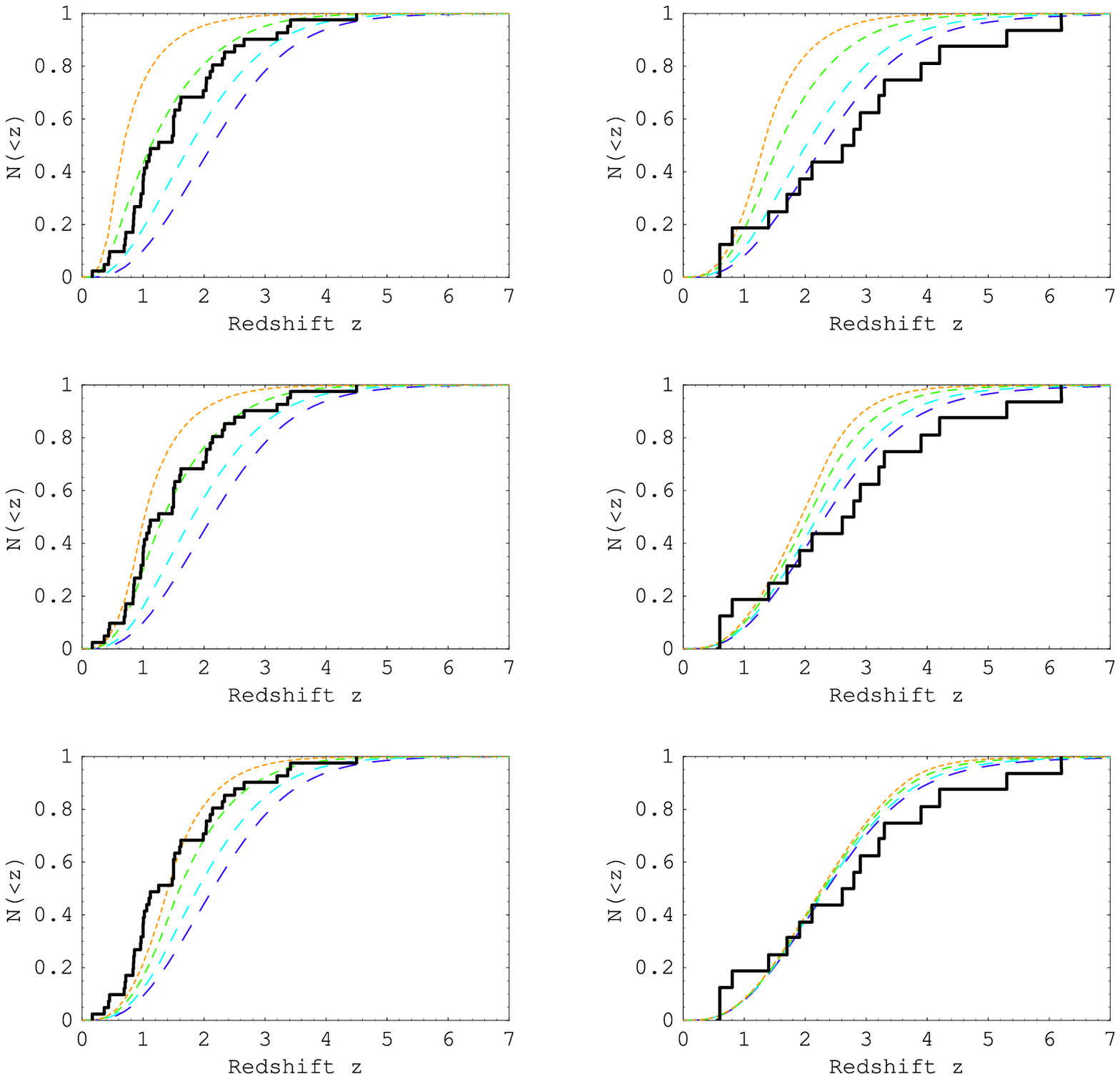}
\vskip-1.8 in

\caption{SFR3: Same as Figure~\ref{fig9} but for
$\Estarg = 5\times10^{51} \ \rm ergs$.}

\label{fig10}
\end{figure}
\clearpage
\begin{figure}[t]
\vskip+.2in \hskip-0.85in
\includegraphics[width=8in]{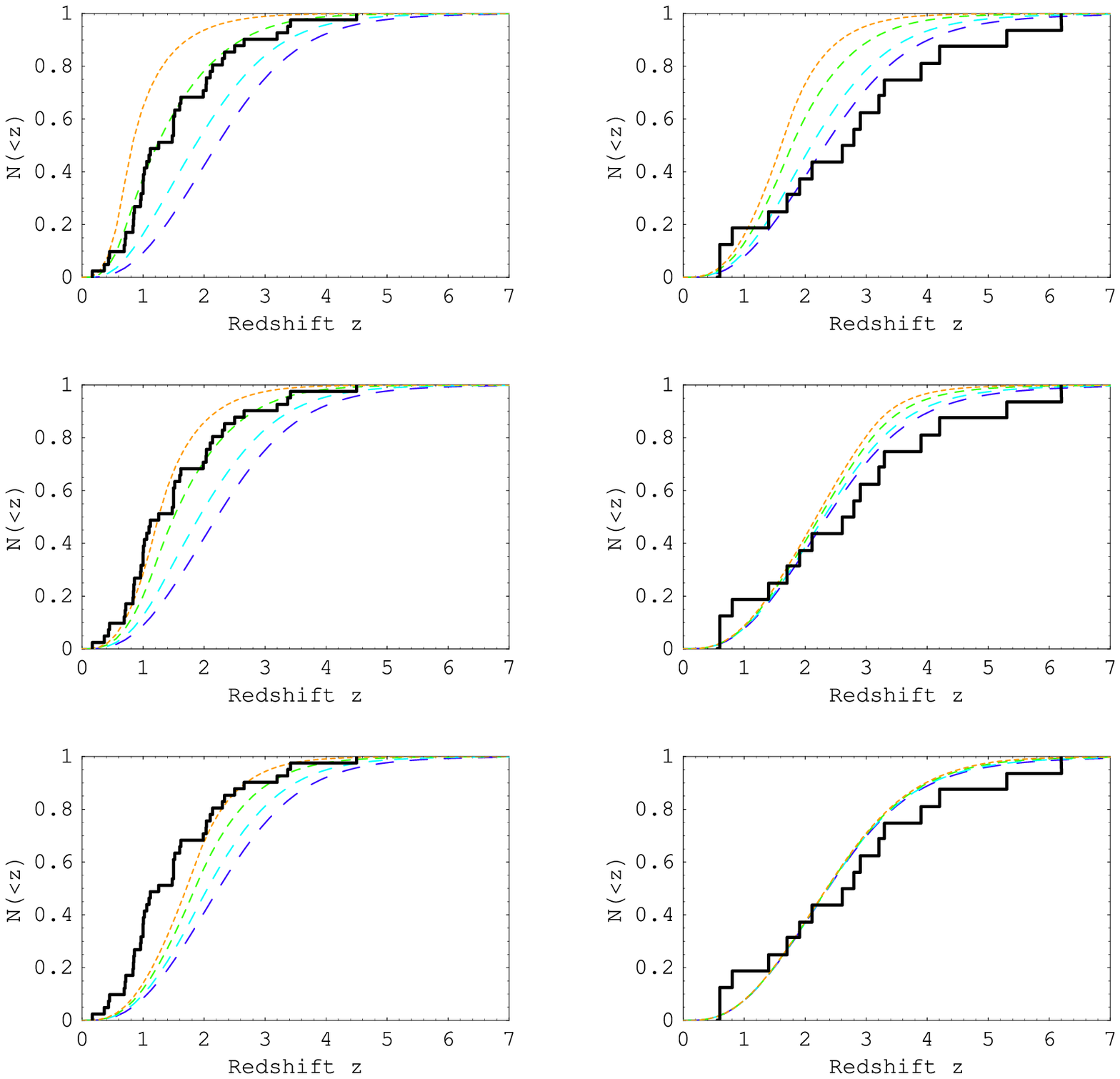}
\vskip-1.8 in

\caption{SFR3: Same as Figure~\ref{fig9} but for
$\Estarg = 10^{52} \ \rm ergs$.}

\label{fig11}
\end{figure}
\clearpage
\begin{figure}[t]
\vskip+.2in \hskip-0.75in
\includegraphics[width=6in]{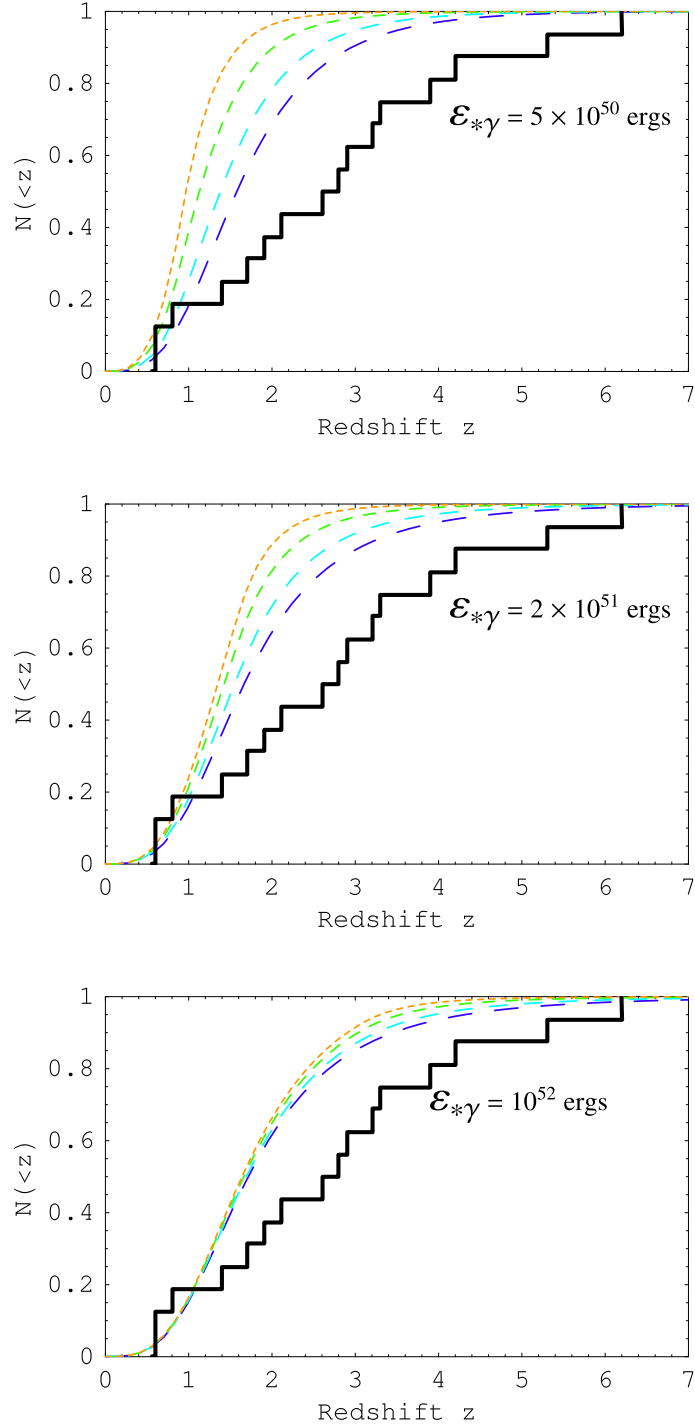}
\vskip-0.6 in

\caption{SFR2: \Swift~cumulative redshift distribution with an
assumed range of the jet opening angles $\theta_{\rm jmin}=0.05$
and $\theta_{\rm jmax}= 0.7 \ \rm radians$. From top to bottom,
the assumed gamma-ray energies are $5 \times 10^{50}$, $2 \times
10^{51}$, and $10^{52} \ \rm ergs$, respectively. Each curve
represents the cumulative redshift distribution for different
power-law indices $s$, where $s=-0.6$, $-1.2$, $-1.8$, and $-2.4$,
from far left to far right, respectively.}

\label{fig12}
\end{figure}
\clearpage
\begin{figure}[t]
\vskip+.20in \hskip-0.85in
\includegraphics[width=8in]{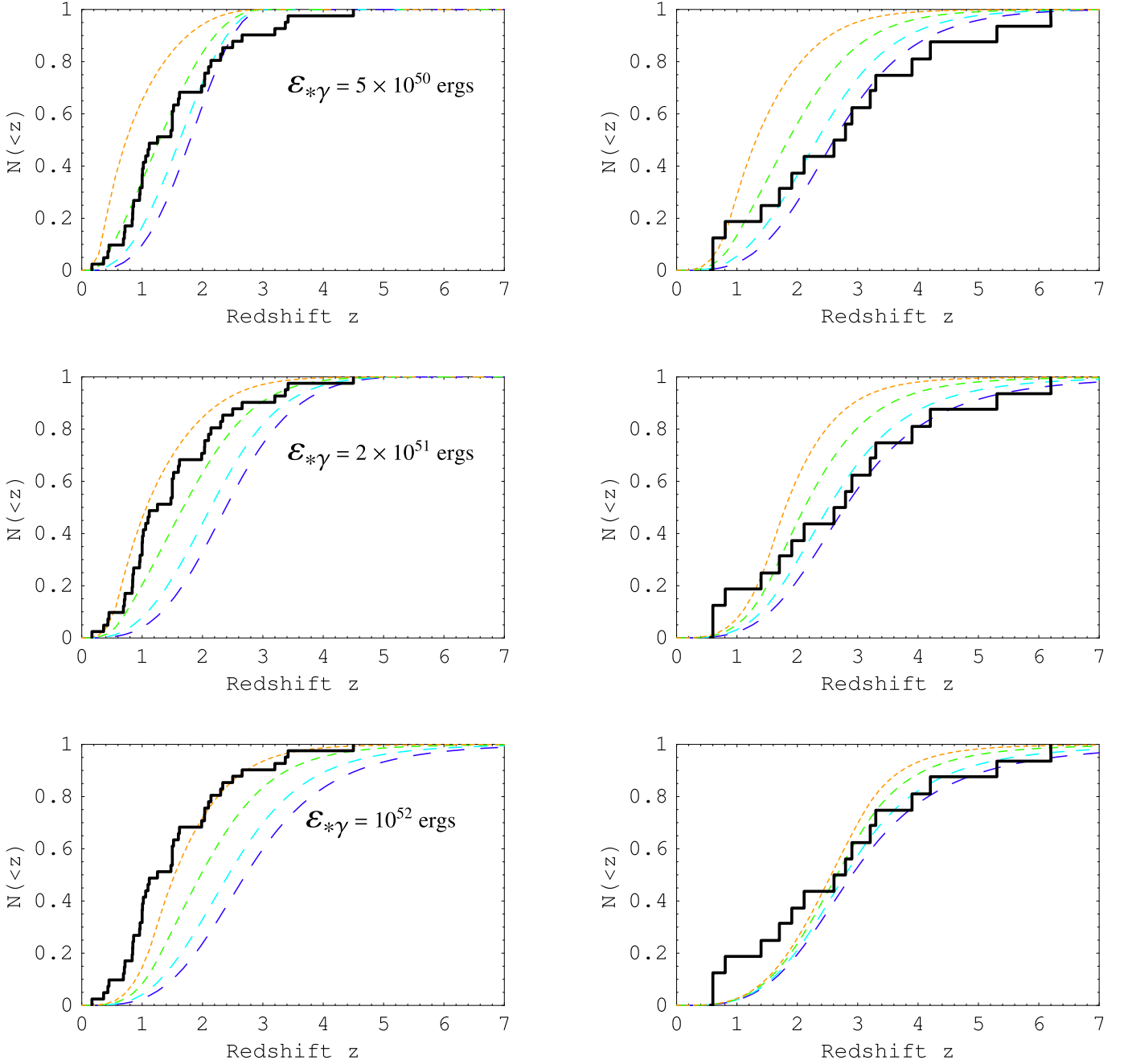}
\vskip-1.8in

\caption{SFR4: In the left column from top to bottom, we have the
pre-\Swift~cumulative redshift distribution with the assumed
gamma-ray energy $\Estarg = 5 \times 10^{50}$, $2 \times 10^{51}$,
and $10^{52} \ \rm ergs$, and an assumed range of the jet opening
angles between $\theta_{\rm jmin}=0.05$ and $\theta_{\rm jmax}=0.7
\ \rm radians$. The left and right columns show the model
distributions for pre-\Swift~and \Swift~data, respectively. In
each panel, each curve represents the cumulative redshift
distribution for different indices $s$, where $s=-0.6$, $-1.2$,
$-1.8$, and $-2.4$, from far left to far right, respectively.}

\label{fig13}
\end{figure}
\clearpage
\begin{figure}[t]
\vskip+0.2in
\includegraphics[width=6in]{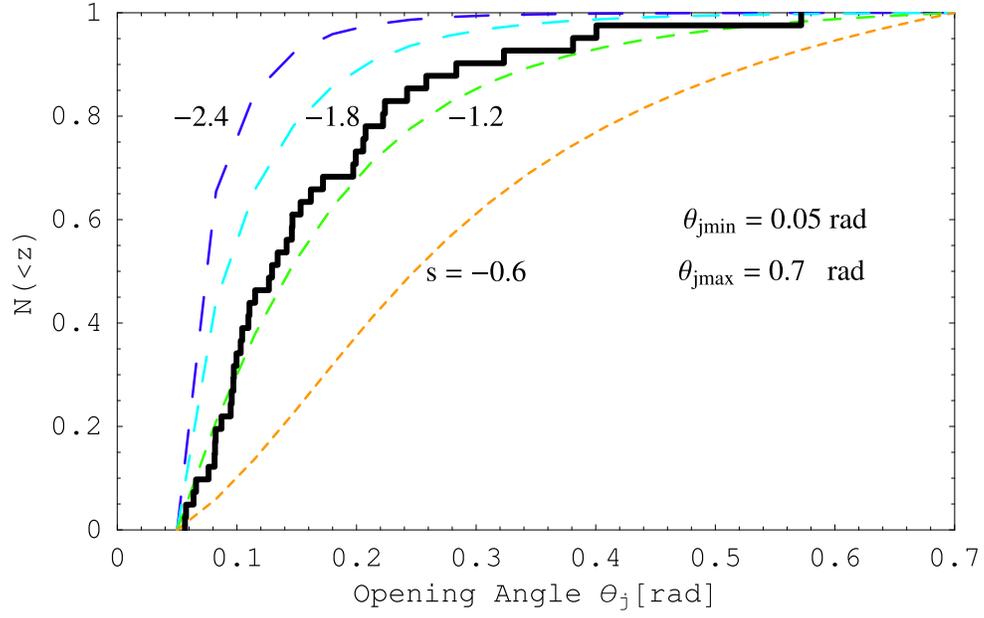}
\vskip-3.45in

\caption{SFR4: pre-\Swift~cumulative opening-angle distribution
with an $\Estarg = 10^{52} \ \rm ergs$, and an assumed range of
the jet opening angles between $\theta_{\rm jmin}=0.05$ and
$\theta_{\rm jmax}=0.7 \ \rm radians$. Each curve represents the
cumulative opening-angle distribution for different indices $s$,
where $s=-0.6$, $-1.2$, $-1.8$, and $-2.4$, from far right to far
left, respectively.}

\label{fig14}
\end{figure}
\clearpage
\begin{figure}[t]
\vskip+0.20in \hskip-0.85in
\includegraphics[width=8in]{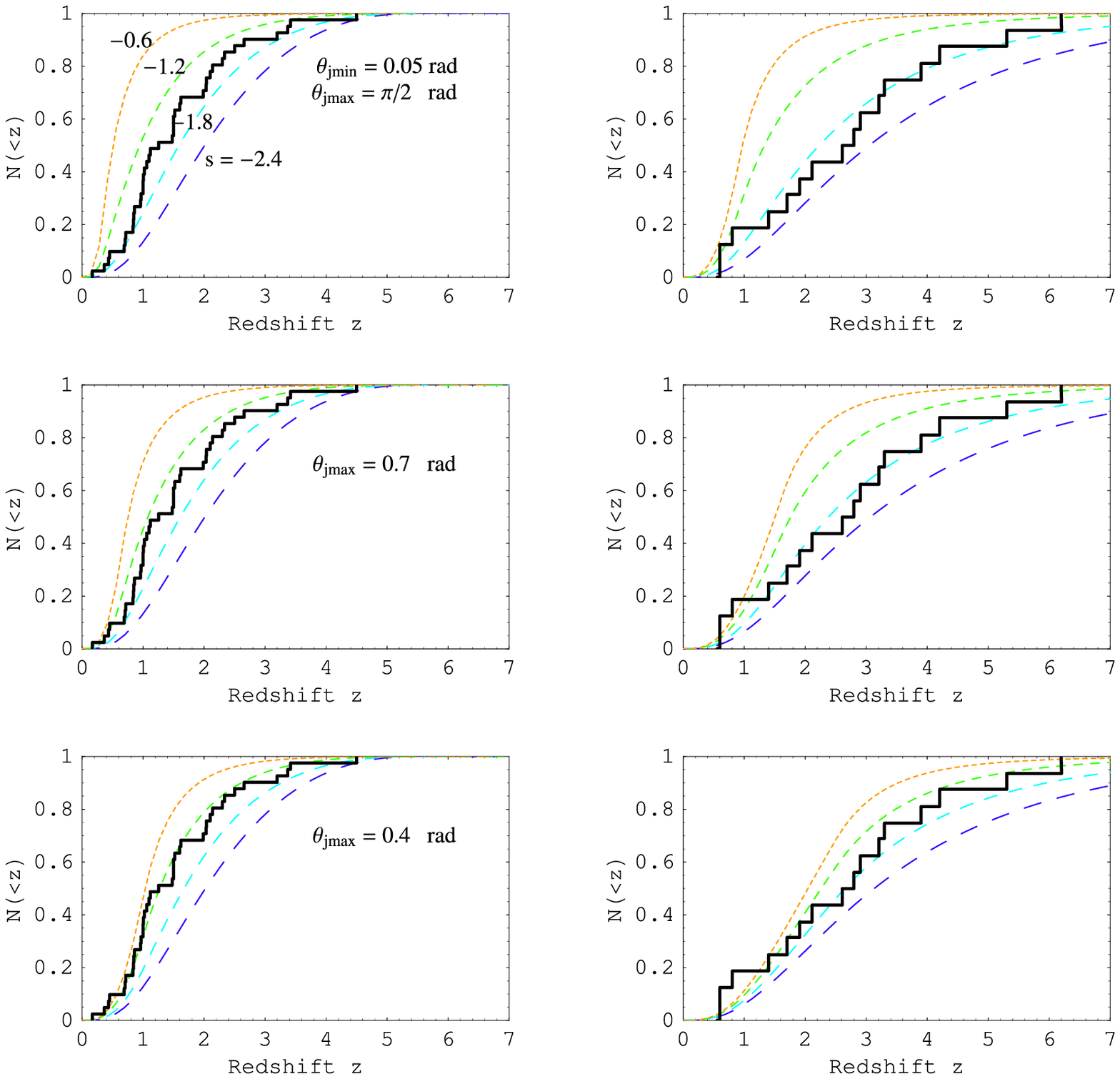}
\vskip-1.8 in

\caption{SFR5: In the left column, from top to bottom, we have the
pre-\Swift~cumulative redshift distribution with the jet opening
angles ranging from $\theta_{\rm jmin}=0.05$ to $\theta_{\rm
jmax}=\pi/2$, $0.7$, and $0.4 \ \rm radians$, respectively. In the
right column we have the same thing as in the left column but for
\Swift~data. In each panel, each curve represents the cumulative
redshift distribution for indices $s$, where $s=-0.6$, $-1.2$,
$-1.8$, and $-2.4$, from far left to far right, respectively. The
assumed gamma-ray energy is $\Estarg = 2 \times 10^{51} \ \rm
ergs$.}

\label{fig15}
\end{figure}
\clearpage
\begin{figure}[t]
\vskip+0.20in \hskip-0.85in
\includegraphics[width=8in]{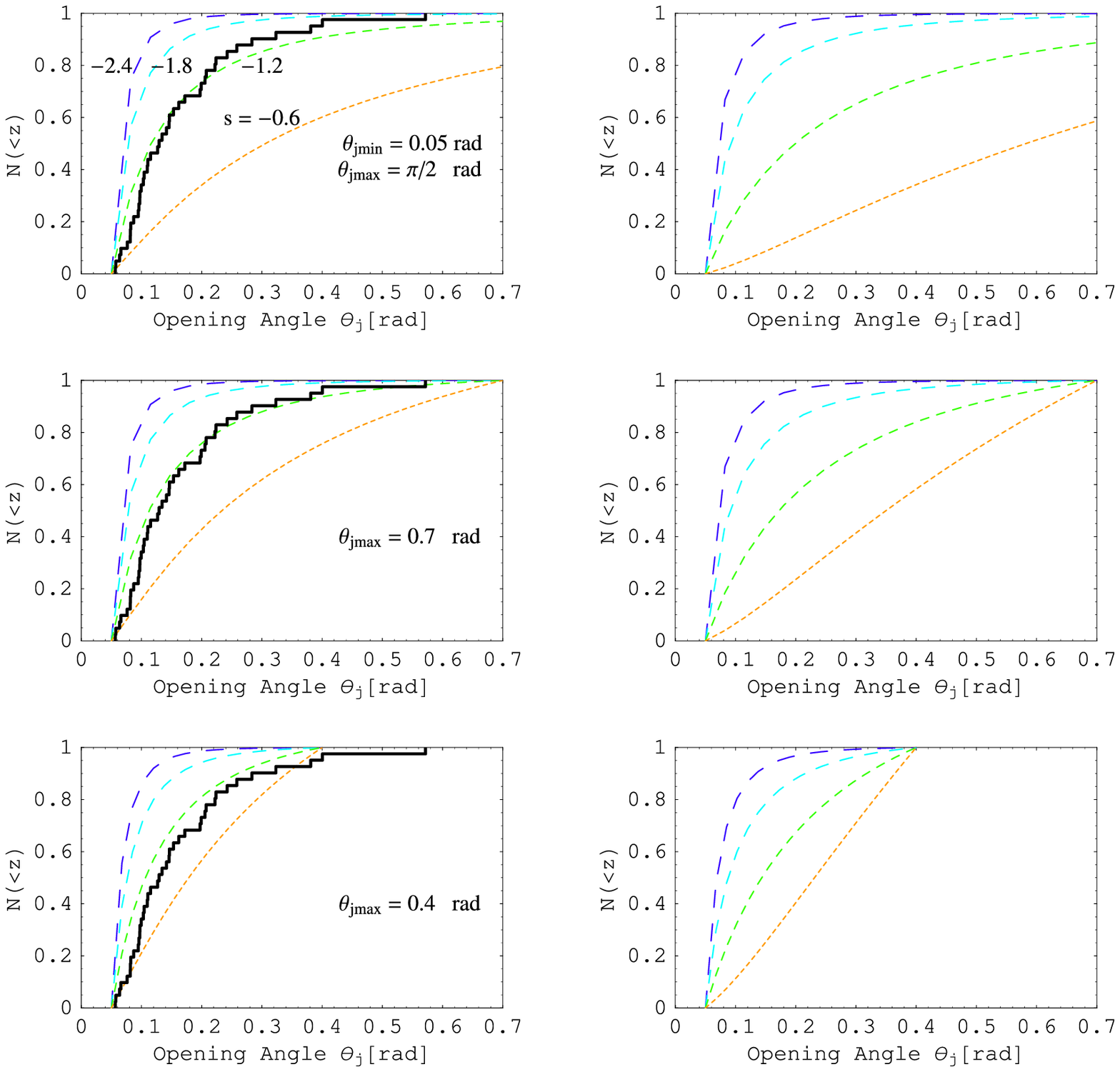}
\vskip-1.8 in

\caption{SFR5: In the left column from top to bottom, we have the
pre-\Swift~cumulative opening-angle distribution with the jet
opening angles ranging from $\theta_{\rm jmin}=0.05$ to
$\theta_{\rm jmax}=\pi/2$, $0.7$, and $0.4 \ \rm radians$,
respectively. In the right column we have the same thing as in the
left column but for the \Swift~data. In each panel, each curve
represents the cumulative redshift distribution for indices
$s=-0.6$, $-1.2$, $-1.8$, and $-2.4$, from far right to far left,
respectively. The assumed gamma-ray energy is $\Estarg = 2 \times
10^{51} \ \rm ergs$.}

\label{fig16}
\end{figure}
\clearpage
\begin{figure}[t]
\vskip+.2in \hskip-0.85in
\includegraphics[width=8in]{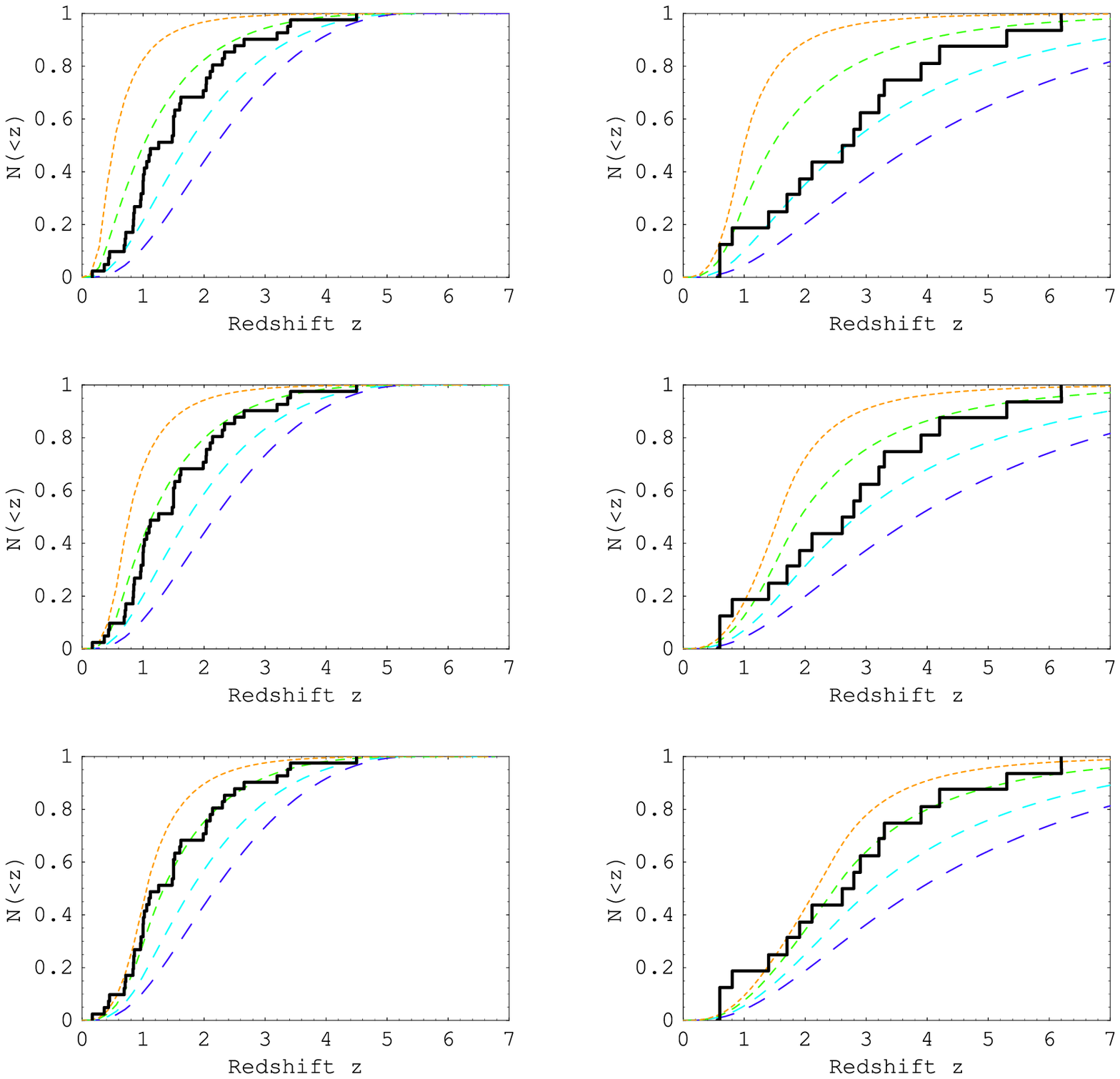}
\vskip-1.8 in

\caption{SFR6: Same as Figure~\ref{fig15} with
$\Estarg =2 \times 10^{51} \ \rm ergs$.}

\label{fig17}
\end{figure}
\clearpage
\begin{figure}[t]
\vskip+.2in \hskip-0.85in
\includegraphics[width=8in]{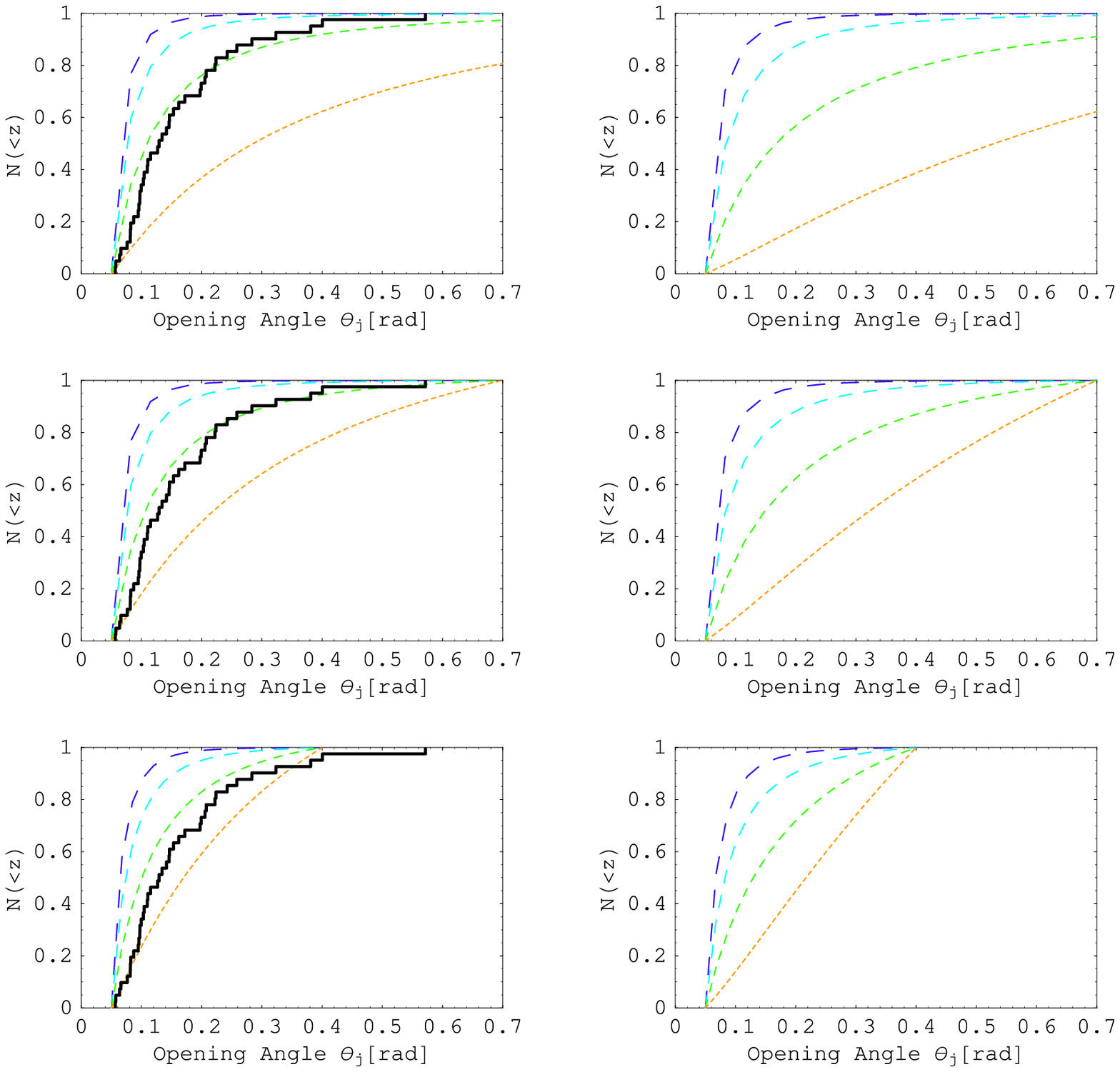}
\vskip-1.8 in

\caption{SFR6: Same as Figure~\ref{fig16} with
$\Estarg =2 \times 10^{51} \ \rm ergs$.}

\label{fig18}
\end{figure}
\clearpage
\begin{figure}[t]
\vskip+0.2in \hskip-0.85in
\includegraphics[width=8in]{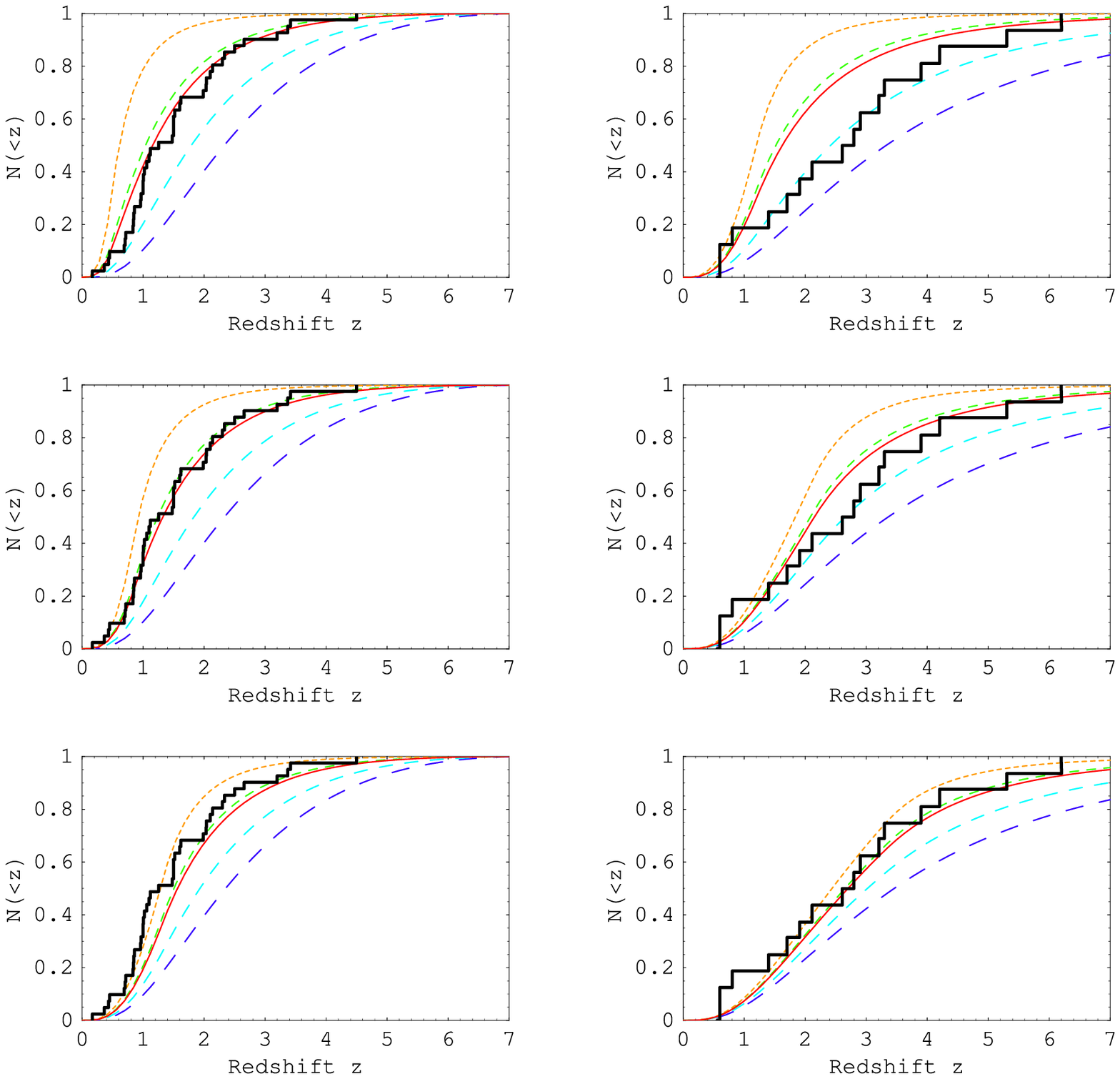}
\vskip-1.8 in

\caption{SFR5: Same as Figure~\ref{fig15} with $\Estarg =4 \times
10^{51} \ \rm ergs$.  The solid curves are for $s=-1.3$.}

\label{fig19}
\end{figure}
\clearpage
\begin{figure}[t]
\vskip+0.2in \hskip-0.85in
\includegraphics[width=8in]{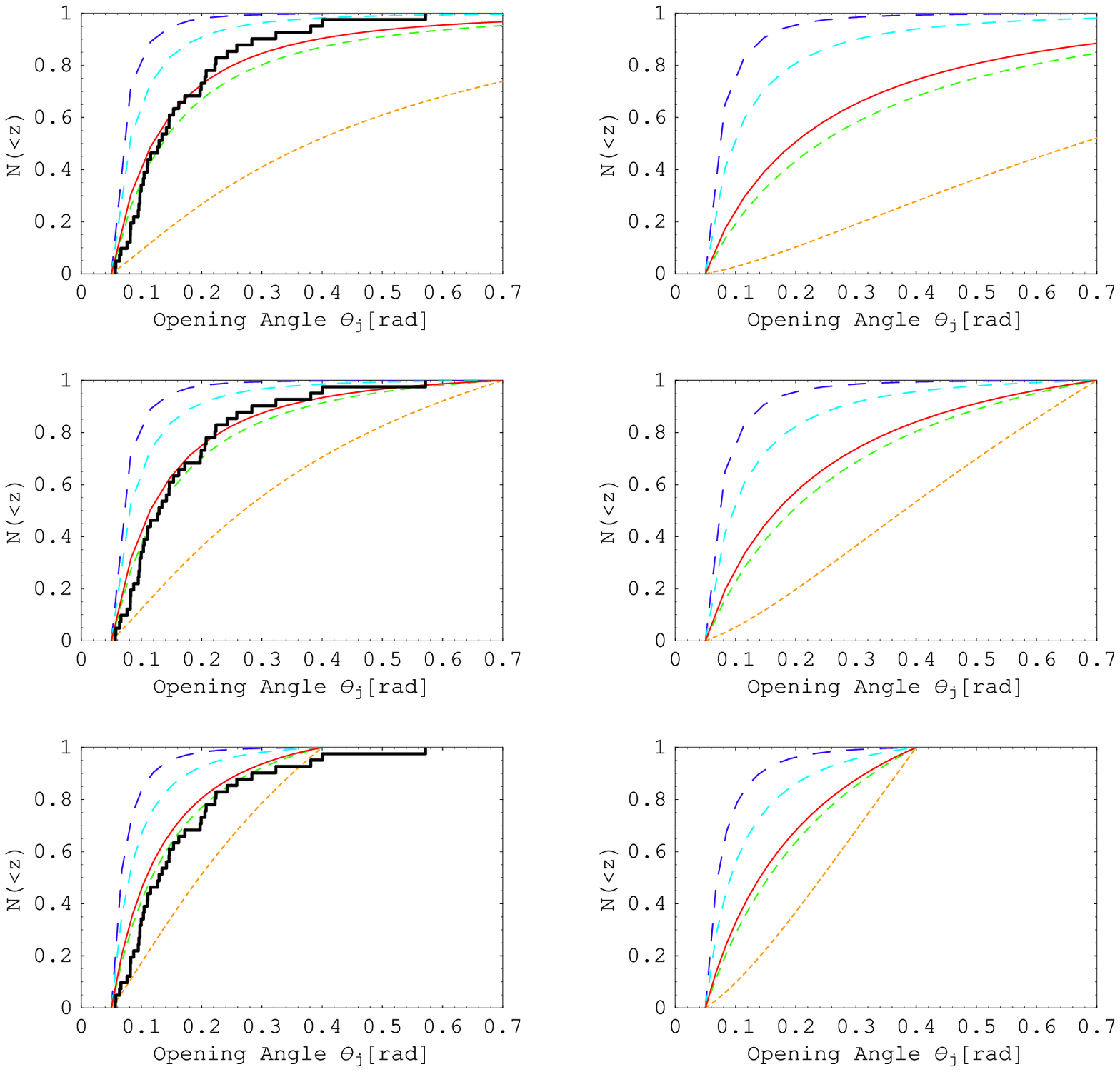}
\vskip-1.8 in

\caption{SFR5: Same as Figure~\ref{fig16} with $\Estarg = 4 \times
10^{51} \ \rm ergs$.  The solid curves are for $s=-1.3$.}

\label{fig20}
\end{figure}
\clearpage
\begin{figure}[t]
\vskip+0.2in \hskip-0.85in
\includegraphics[width=8in]{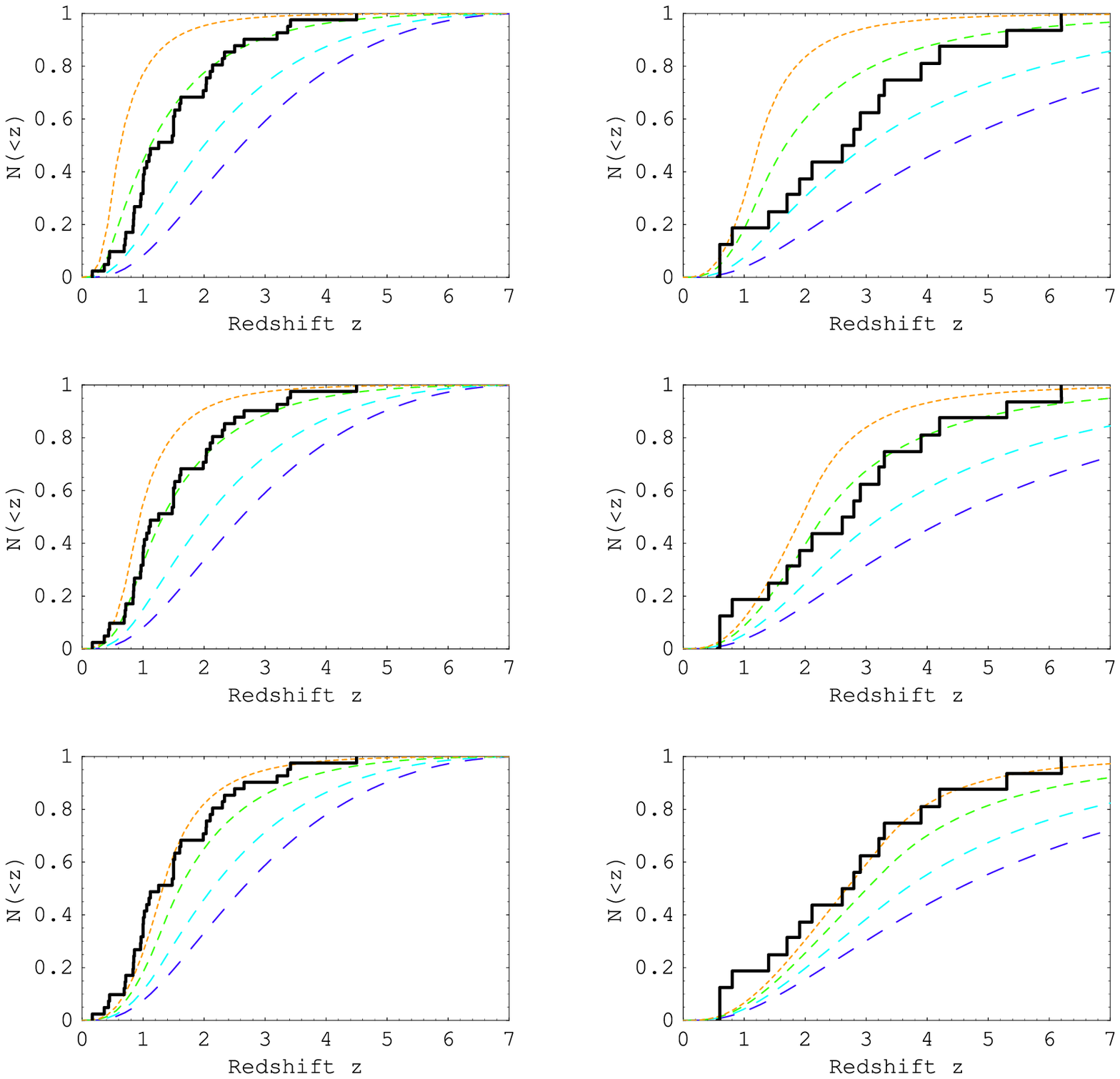}
\vskip-1.8 in

\caption{SFR6: Same as Figure~\ref{fig15} with
$\Estarg = 4 \times 10^{51} \ \rm ergs$.}

\label{fig21}
\end{figure}
\clearpage
\begin{figure}[t]
\vskip+0.2in \hskip-0.85in
\includegraphics[width=8in]{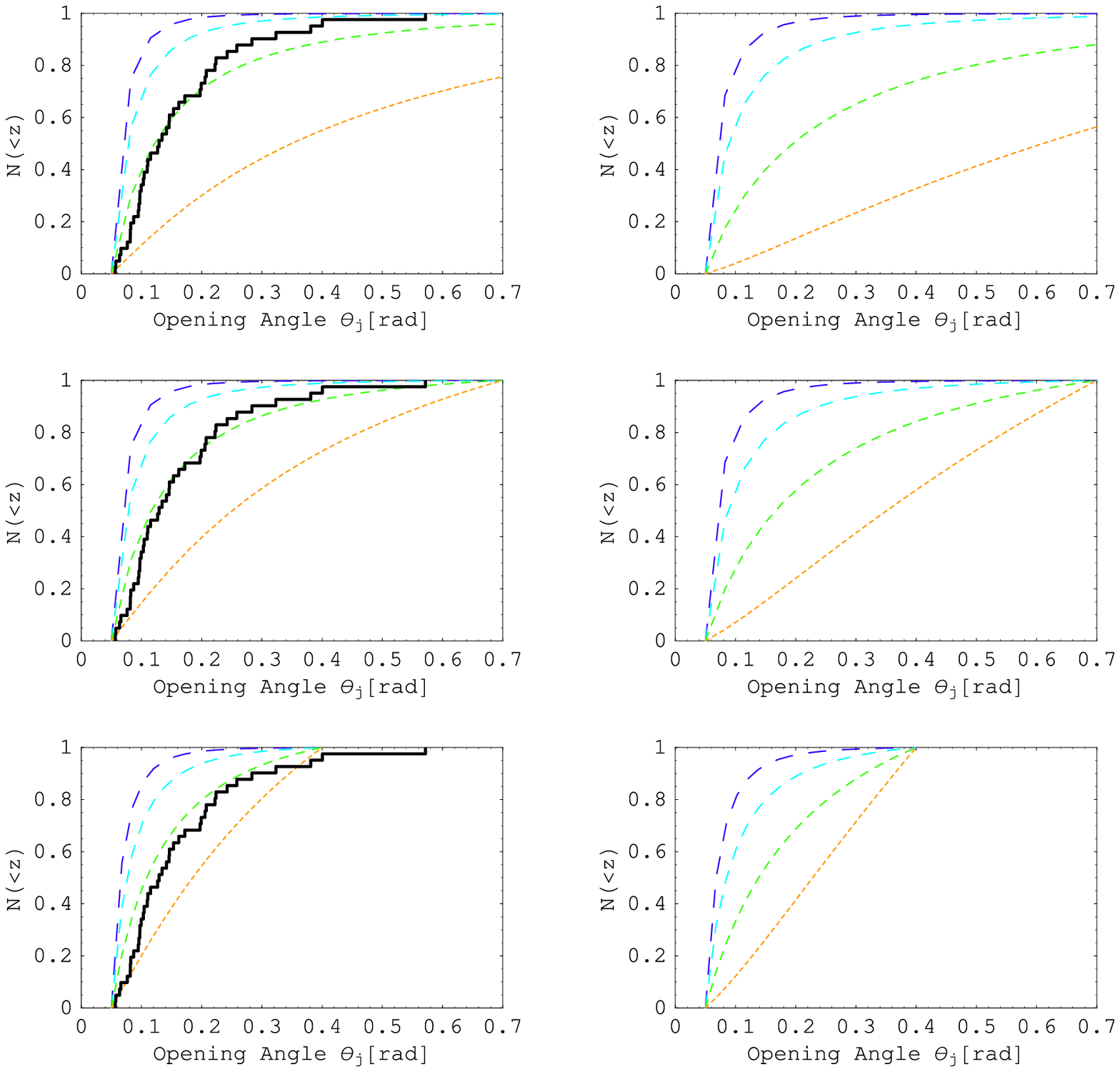}
\vskip-1.8 in

\caption{SFR6: Same as Figure~\ref{fig16} with
$\Estarg = 4 \times 10^{51} \ \rm ergs$.}

\label{fig22}
\end{figure}
\clearpage
\begin{figure}[t]
\vskip+0.2in \hskip-0.85in
\includegraphics[width=8in]{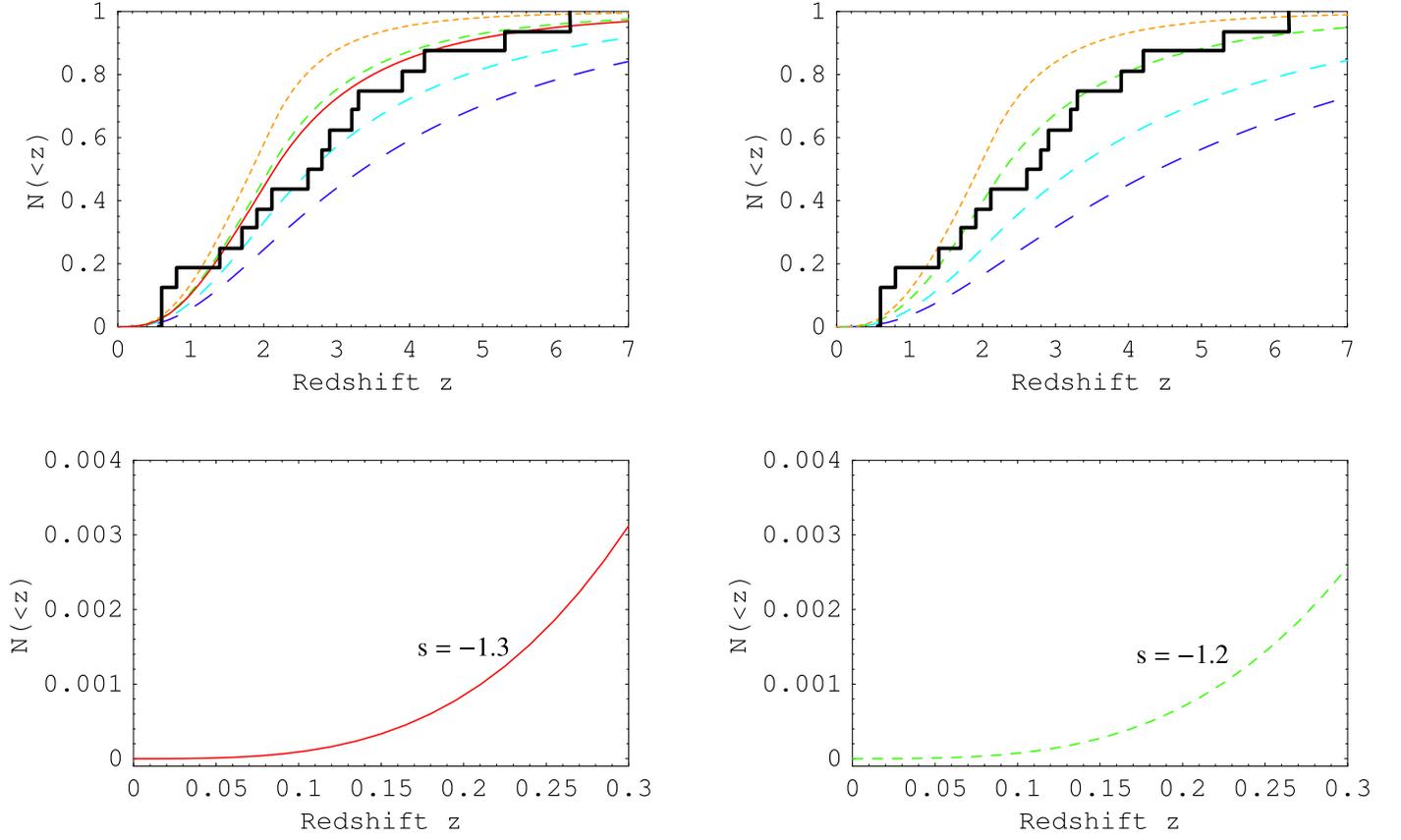}
\vskip-3. in

\caption{\Swift~cumulative redshift distribution assuming SFR5
(left column) and SFR6 (right column), with $\Estarg = 4 \times
10^{51} \ \rm ergs$. The two bottom panels show the above
best-fitted results at $z \leq 1$ in more detail.}

\label{fig23}
\end{figure}
\clearpage
\begin{figure}[t]
\includegraphics[width=6in]{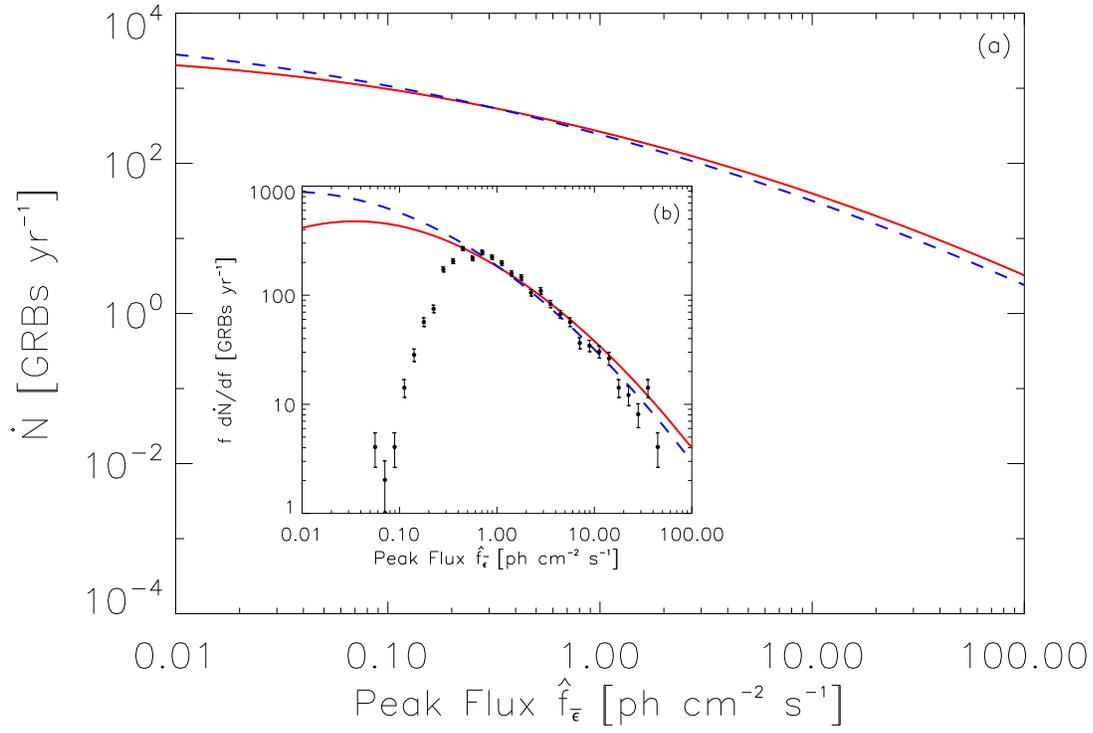}

\caption{Integral size distributions (a) for SFR5 (solid) and SFR6
(dash curve), respectively. The inset is the differential size
distributions (b) for SFR5 (solid) and SFR6 (dash curve),
respectively. The filled circles curve represents the 1024 ms
trigger timescale from the 4B catalog data, containing 1292
bursts.}

\label{fig24}
\end{figure}
\clearpage

{}


\begin{thebibliography}{}

\bibitem[Amati et al.(2002)]{ama02} Amati, L., et al.\ 2002,
\aap, 390, 81

\bibitem[Bagoly et al.(2006)]{bag06} Bagoly, Z., et al.\
2006, \aap, 453, 797

\bibitem[Band(2002)]{ban02}Band, D.~L.\ 2002, \apj, 578, 806

\bibitem[Band(2003)]{ban03}Band, D.\ L.\  2003, \apj, 588, 945

\bibitem[Band(2006)]{ban06} Band, D.~L.\ 2006, \apj, 644, 378

\bibitem[Barthelmy et al.(2005)] {bar05} Barthelmy, S. D., et al. 2005, Nature, 438, 994

\bibitem[Berger et al.(2006)]{ber06} Berger, E., Fox, D.\ B.,
Kulkarni, S.\ R., Frail, D.~A., \& Djorgovski, S.\ G.\ 2006, \apj,
submitted (astro-ph/0609170)

\bibitem[Berger et al.(2003)]{ber03} Berger, E., Kulkarni,
S.~R., Frail, D.~A., \& Soderberg, A.~M.\ 2003, \apj, 599, 408

\bibitem[Berger et al.(2005)]{ber05}Berger, E. et al.  2005, \apj, 634, 501

\bibitem[Blain \& Natarajan(2000)]{bn00}Blain, A. W., \& Natarajan, P. 2000, \apj, 312, L35

\bibitem[Blain et al.(1999)]{bla99}Blain, A. W. et al.  1999, \mnras, 309, 715

\bibitem[Bloom et al.(2003)]{blo03}Bloom, J. S., Frail, D. A.,
\& Kulkarni, S. R.  2003, \apj, 594, 674

\bibitem[Borgonovo(2004)]{bor04} Borgonovo, L.\ 2004, \aap,
418, 487

\bibitem[B\"ottcher \& Dermer(2000)]{bd00}B\"ottcher, M.,
\& Dermer, C. D.  2000, \apj, 529, 635

\bibitem[Bromm \& Loeb(2002)]{bl02} Bromm, V., \& Loeb, A.\
2002, \apj, 575, 111

\bibitem[Bromm \& Loeb(2006)]{bl06} Bromm, V., \& Loeb, A.\
2006, \apj, 642, 382

\bibitem[Cappellaro et al.(1999)] {cap99} Cappellaro, E., Evans, R.,
\& Turatto, M. 1999, \aap, 351, 459

\bibitem[Ciardi \& Loeb(2000)]{cl00} Ciardi, B., \& Loeb,
A.\ 2000, \apj, 540, 687

\bibitem[Daigne et al.(2006)]{drm06} Daigne, F., Rossi, E.,
\& Mochkovitch, R.\ 2006, \mnras, 372, 1034

\bibitem[Della Valle et al.(2006)]{val06} DellaValle, M., et al.\
2006, Nature, 444, 1050

\bibitem[Dermer(1992)]{der92}Dermer, C. D. 1992, Phys. Rev. Lett., 68, 1799

\bibitem[Dermer(2006)]{der06}Dermer, C. D. 2006, \apj, in press (astro-ph/0605402)

\bibitem[Dermer \& Holmes(2005)]{dh05} Dermer, C.~D., \&
Holmes, J.~M.\ 2005, \apjl, 628, L21

\bibitem[Frail et al.(2001)]{fra01}Frail, D. A. et al. 2001, \apj, 562, L55

\bibitem[Friedman \& Bloom(2005)]{fb05} Friedman, A.~S., \& Bloom, J.~S.
2005, \apj, 627, 1

\bibitem[Fruchter et al.(2006)]{fru06} Fruchter, A.~S., et
al.\ 2006, \nat, 441, 463

\bibitem[Fynbo et al.(2006)]{fyn06} Fynbo, J.~P.~U., et al.\
2006, Nature, 444, 1047

\bibitem[Fynbo et al.(2006a)]{fyn06a} Fynbo, J.~P.~U., et al.\
2006a, \aap, 451, L47

\bibitem[Gal-Yam et al.(2006)]{gal06} Gal-Yam et al., 2006, Nature, 444, 1053

\bibitem[Gehrels et al.(2004)]{geh04}Gehrels, N. et al. 2004, \apj, 611, 1005

\bibitem[Gehrels et al.(2006)]{geh06}Gehrels, N. et al. 2006, Nature, 444, 1044

\bibitem[Ghirlanda et al.(2004)]{ghi04}Ghirlanda, G. et al. 2004, \apj, 613, L13

\bibitem[Guetta, Piran, \& Waxman(2005)]{gpw05}Guetta, D., Piran, T.,
\& Waxman, E. 2005, \apj, 619, 412

\bibitem[Hopkins \& Beacom(2006)]{hb06}Hopkins, A. M., \& Beacom, J. F. 2006, \apj,
651, 142

\bibitem[Horv{\'a}th et al.(2006)]{hor06} Horv{\'a}th, I.,
Bal{\'a}zs, L.~G., Bagoly, Z., Ryde, F., \& M{\'e}sz{\'a}ros, A.\ 2006,
\aap, 447, 23

\bibitem[Jakobsson et al.(2006)]{jak06}Jakobsson, P. et al. 2006, A\&A, 447, 897

\bibitem[Lamb \& Reichart(2000)]{lr00}Lamb, D. Q., \& Reichart, D. E.  2000, \apj, 536, 1

\bibitem[Liang et al.(2006)]{lia06} Liang, E., Zhang, B., Virgili, F., \& Dai, Z.\ G.\ 2006,
\apj, submitted (astro-ph/0605200)

\bibitem[Lloyd-Ronning, Fryer, \& Ramirez-Ruiz(2002)]{lfr02} Lloyd-Ronning,
N.~M., Fryer, C.~L., \& Ramirez-Ruiz, E.\ 2002, \apj, 574, 554

\bibitem[Loveday et al.(1992)]{lov92} Loveday, J., Peterson,
B.~A., Efstathiou, G., \& Maddox, S.~J.\ 1992, \apj, 390, 338

\bibitem[Melott et al.(2004)]{mel04} Melott, A.~L., et al.\
2004, International Journal of Astrobiology, 3, 55

\bibitem[M{\'e}sz{\'a}ros et al.(2006)]{mbbh06} M{\'e}sz{\'a}ros, A., Bagoly, Z.,
Balazs,  L.\ G., \& Horvath, I., 2006, A\&A, 455 785

\bibitem[M{\'e}sz{\'a}ros \& M{\'e}sz{\'a}ros(1996)]{mm96} M{\'e}sz{\'a}ros, A., \&
M{\'e}sz{\'a}ros, P.\ 1996, \apj, 466, 29

\bibitem[M{\'e}sz{\'a}ros(2006)]{mes06} M{\'e}sz{\'a}ros, P.\
2006, Reports of Progress in Physics, 69, 2259

\bibitem[M{\'e}sz{\'a}ros \& Rees(1997)]{mr97} M{\'e}sz{\'a}ros, P., \&
Rees, M.~J.\ 1997, \apjl, 482, L29

\bibitem[M{\'e}sz{\'a}ros, Rees \& Wijers(1998)]{mrw98}M{\'e}sz{\'a}ros, P.,
Rees, M. J., \& Wijers, R. A. M. J. 1998, \apj, 499, 301

\bibitem[Mortsell \& Sollerman(2005)]{ms05}Mortsell, E., \& Sollerman, J.  2005, JCAP, 6, 9

\bibitem[Natarajan et al.(2005)]{nat05} Natarajan, P.,
Albanna, B., Hjorth, J., Ramirez-Ruiz, E., Tanvir, N., \& Wijers, R.\ 2005,
\mnras, 364, L8

\bibitem[Paciesas et al.(1999)]{pac99} Paciesas, W.~S., et
al.\ 1999, \apjs, 122, 465

\bibitem[Panaitescu \& Kumar(2001)]{pk01}Panaitescu, A., \& Kumar, P.  2001, \apj, 560, L49

\bibitem[Peebles(1993)]{pee93} Peebles, P.~J.~E.\ 1993,
Princeton Series in Physics, Princeton, NJ: Princeton University Press

\bibitem[Perna, Sari, \& Frail(2003)]{psf03} Perna, R., Sari, R., \&
Frail, D.\ 2003, \apj, 594, 379

\bibitem[Porciani \& Madau(2001)]{pm01} Porciani, C., \&
Madau, P.\ 2001, \apj, 548, 522

\bibitem[Rossi, Lazzati \& Rees(2002)]{rlr02}Rossi, E., Lazzati, D., \& Rees, M. J.
2002, \apj, 332, 945

\bibitem[Schmidt(1999)]{sch99} Schmidt, M.\ 1999, \apjl, 523,
L117

\bibitem[Spergel et al.(2003)]{spe03} Spergel, D.~N., et al.\ 2003, \apjs, 148, 175

\bibitem[Stanek et al.(1999)]{sta99} Stanek, K.~Z.,
Garnavich, P.~M., Kaluzny, J., Pych, W., \& Thompson, I.\ 1999, \apjl, 522,
L39

\bibitem[Stanek et al.(2006)]{sta06} Stanek, K.~Z. et al. 2007, Acta Astronomica, in press, (astro-ph/0604113)

\bibitem[Tavani(1998)]{tav98} Tavani, M.\ 1998, \apjl, 497, L21

\bibitem[Thomas et al.(2005)]{tho05} Thomas, B.~C., Jackman,
C.~H., Melott, A.~L., Laird, C.~M., Stolarski, R.~S., Gehrels, N.,
Cannizzo, J.~K., \& Hogan, D.~P.\ 2005, \apjl, 622, L153

\bibitem[Totani(1997)]{tot97} Totani, T.\ 1997, \apjl, 486,
L71
\bibitem[Totani(1999)]{tot99}Totani, T.  1999, \apj, 511, 41

\bibitem[Waxman, Kulkarni \& Frail(1998)]{wkf98}Waxman, E., Kulkarni, S. R.,
\& Frail, D. A  1998, \apj, 497, 288

\bibitem[Wick, Dermer \&  Atoyan(2004)]{wda04}Wick, S. D., Dermer, C. D.,
\& Atoyan, A.  2004, Astropart. Phys., 21, 125

\bibitem[Wijers et al.(1998)]{wij98} Wijers, R.~A.~M.~J.,
Bloom, J.~S., Bagla, J.~S., \& Natarajan, P.\ 1998, \mnras, 294, L13

\bibitem[Yamazaki et al.(2003)]{yam03} Yamazaki, R.,
Yonetoku, D., \& Nakamura, T.\ 2003, \apjl, 594, L79

\bibitem[Zhang(2006)]{zha06} Zhang, B. \ 2006, Nature, 444, 1010

\bibitem[Zhang et al.(2004)]{zha04} Zhang, B., Dai, X.,
Lloyd-Ronning, N.~M., \& M{\'e}sz{\'a}ros, P.\ 2004, \apjl, 601, L119

\end{thebibliography}
\end{document}